\newif\ifsubmode
\submodefalse

\newif\ifprintfig
\printfigtrue

\newif\ifemulate
\emulatetrue

\ifsubmode
  \emulatefalse
  \documentclass[12pt,preprint]{aastex}
  \usepackage{epsfig}
  \received{}
  \accepted{}
  \journalid{}{}
  \articleid{}{}
\else
  \ifemulate
    \documentclass[apj]{emulateapj}
  \else
    \documentclass[preprint]{aastex}
  \fi
  \usepackage{epsfig}
  \slugcomment{{\it ApJ, submitted 11/17/2006}}
\fi

\shorttitle{Dynamical Models of Elliptical Galaxies in $z = 0.5$ clusters: I} 
\shortauthors{van der Marel \& van Dokkum}

\newcommand{\etal}{{et al.~}}
\newcommand{\lta}{\lesssim}
\newcommand{\gta}{\gtrsim}
\newcommand{\kms}{\>{\rm km}\,{\rm s}^{-1}}

\newcommand{\pc}{\>{\rm pc}}
\newcommand{\kpc}{\>{\rm kpc}}
\newcommand{\Mpc}{\>{\rm Mpc}}
\newcommand{\Msun}{\>{\rm M_{\odot}}}
\newcommand{\Lsun}{\>{\rm L_{\odot}}}

\begin{document}

\title{Dynamical Models of Elliptical Galaxies in $z = 0.5$ Clusters:\\
I.~Data-Model Comparison and Evolution of Galaxy Rotation}

\author{Roeland P.~van der Marel}
\affil{Space Telescope Science Institute, 3700 San Martin Drive, 
       Baltimore, MD 21218}

\author{Pieter G.~van Dokkum}
\affil{Department of Astronomy, Yale University, New Haven, CT 06520}

%%%%%%%%%%%%%%%
% Use a small baselineskip, unless in submission mode.
%%%%%%%%%%%%%%%

\ifsubmode\else\ifemulate\else
\baselineskip=14pt
\fi\fi

%%%%%%%%%%%%%%%
% Abstract
%%%%%%%%%%%%%%%

\begin{abstract}
We present spatially resolved stellar rotation velocity and velocity
dispersion profiles form Keck/LRIS absorption-line spectra for 25
galaxies, mostly visually classified ellipticals, in three clusters at
$z \approx 0.5$. We interpret the kinematical data and HST photometry
using oblate axisymmetric two-integral $f(E,L_z)$ dynamical models
based on the Jeans equations. This yields good fits, provided that the
seeing and observational characteristics are carefully modeled. The
fits yield for each galaxy the dynamical $M/L$ and a measure of the
galaxy rotation rate. Paper~II addresses the implied $M/L$
evolution. Here we study the rotation-rate evolution by comparison to
a sample of local elliptical galaxies of similar present-day
luminosity. The brightest galaxies in the sample all rotate too slowly
to account for their flattening, as is also observed at $z=0$. But the
average rotation rate is higher at $z \approx 0.5$ than locally. This
may be due to a higher fraction of misclassified S0 galaxies (although
this effect is insufficient to explain the observed strong evolution
of the cluster S0 fraction with redshift). Alternatively, dry mergers
between early-type galaxies may have decreased the average rotation
rate over time. It is unclear whether such mergers are numerous enough
in clusters to explain the observed trend quantitatively. Disk-disk
mergers may affect the comparison through the so-called ``progenitor
bias'', but this cannot explain the direction of the observed
rotation-rate evolution. Additional samples are needed to constrain
possible environmental dependencies and cosmic variance in galaxy
rotation rates. Either way, studies of the internal stellar dynamics
of distant galaxies provide a valuable new approach for exploring
galaxy evolution.
\end{abstract}

%%%%%%%%%%%%%%%
% Keywords
%%%%%%%%%%%%%%%


%%%%%%%%%%%%%%%
% Beginning of main text
%%%%%%%%%%%%%%%

\section{Introduction}
\label{s:intro}

Elliptical galaxies form an important component of the galaxy
population in the universe. In fact, they constitute the bulk of the
population both at the high-mass end of the galaxy distribution and at
the highest environmental densities (i.e., in clusters). An
understanding of the formation and evolution of elliptical galaxies is
therefore crucial for any successful theory of hierarchical structure
formation in a cosmological context.

The most direct way to constrain the evolution of elliptical galaxies
is through observations of their photometric and dynamical properties
at intermediate and high redshifts. Several of the most successful
techniques to use such data have relied on the statistical properties
of globally averaged or characteristic galaxy quantities. For example,
the fundamental plane is a correlation between the effective radius
$r_{\rm eff}$, the average surface brightness $I_{\rm eff}$ inside the
effective radius, and the velocity dispersion $\sigma$. The evolution
of the fundamental plane can be used to place constraints on the
formation redshifts of elliptical galaxies (as reviewed in van Dokkum
\& van der Marel 2006, hereafter vDvdM06). Similarly, the 
color-magnitude relation is a correlation between the total magnitude
and globally averaged color. Elliptical galaxies form a ``red
sequence'' that can also be used to constrain the formation redshifts
of elliptical galaxies (e.g., De Lucia \etal 2004; Mei \etal 2006).

The success and popularity of studies based entirely on globally
averaged or characteristic galaxy properties can be attributed at
least in part to the relative ease with which such studies can be
performed. In particular, no modeling or understanding of the internal
structure of the sample galaxies is necessarily required. However,
this is at the same time the main drawback of such studies. To draw
conclusions one is usually forced to make some implicit assumptions
about the structure of the galaxy, or at least, that this structure
does not evolve with time.  Such assumptions remain untested until one
actually progresses to more detailed studies of the resolved galaxy
structure. In fact, data that enable such studies are often available,
but remain unused. For example, fundamental plane or color-magnitude
relation studies do not use available information on the axial ratios
of individual galaxies, their resolved surface brightness or color
profiles, or any spatially resolved information on the kinematical
structure of the galaxy.

One approach to learn more about the internal structure of distant
elliptical galaxies is to construct detailed dynamical models for
high-quality spatially resolved photometric and kinematic data. The
required dynamical modeling tools exist, and high quality photometric
data is available for many galaxies from the Hubble Space Telescope
(HST). The most challenging part is therefore to obtain the necessary
kinematic data. However, this is quite possible with large aperture
telescopes equipped with multi-slit spectrographs (e.g., Kelson \etal
2000). Treu \& Koopmans (2004) and Koopmans \etal (2006) obtained and
modeled resolved kinematical profiles for some lensing galaxies. They
constructed relatively simple dynamical models (namely, parameterized
spherical models) but this still allowed them to obtain several
interesting new results on the dark halo component of these
galaxies. There is no reason why similar methodologies or more
detailed modeling approaches cannot be applied to other large samples
of (non-lensing) early-type galaxies.

In vDvdM06 we presented spectroscopy with the Low Resolution Imager
and Spectrograph (LRIS) on Keck of some two dozen galaxies in the
intermediate-redshift ($z \approx 0.5$) clusters CL3C295, CL0016+16
and CL1601+42. The sample galaxies were selected to be bright enough
for spectroscopy, and visually classified from the HST images of
Dressler \etal (1997) and Smail \etal (1997) as early-type (and in
most cases elliptical) galaxies. In vDvdM06 we only used the data for
a study of the FP evolution of the three sample clusters. However, we
obtained spectroscopic data that are deep enough to extract spatially
resolved rotation curves and velocity dispersion profiles.
Furthermore, in the multi-slit mask designs care was taken to tilt
individual slits so that they were aligned preferentially with the
major axes of the galaxies, insofar possible given geometric
restrictions. We present the kinematical profiles here, and we
construct detailed dynamical models to interpret them.

Our models are axisymmetric with two-integral distribution functions
of the form $f=f(E,L_z)$. Here $E$ is the energy and $L_z$ the angular
momentum around the symmetry axis. The internal dynamics are
calculated by solving the Jeans equations of hydrostatic
equilibrium. Models of this type have been constructed successfully
for many nearby galaxies (e.g., Binney, Davies, \& Illingworth 1990;
van der Marel 1991; Magorrian \etal 1998; Corsini \etal
1999). Two-integral models are relatively simple in that they have
$\overline{v_R^2} \equiv \overline{v_z^2}$. So they can be thought of
as the axisymmetric generalization of spherical isotropic models. Of
course, more sophisticated modeling approaches do exist (e.g., those
that built completely general three-integral distribution functions
using the technique of numerical orbit superposition; van der Marel
\etal 1998; Gebhardt \etal 2003). However, the complexity of these
models would be ``overkill'' as compared to the quality of the data
that can be realistically obtained at intermediate redshifts.

The models yield the rotation rate of each galaxy by fitting to the
spatially resolved kinematical profiles. The results can therefore be
used to study the evolution of the rotation rate of galaxies as a
function of time. This evolution can yield new insights into the
mechanisms by which the internal structure of elliptical galaxies gets
modified through mergers. Mergers are believed to be characteristic of
the hierarchical build up of all structure in the Universe. However,
the relative importance of different types of mergers remains poorly
constrained. While the formation of elliptical galaxies through
disk-disk mergers has received considerable attention historically
(e.g., Barnes 1988), much recent work has focused on the importance of
``dry'' mergers between early-type galaxies (van Dokkum 2005; Bell
\etal 2006). Such mergers produce remnants with different rotational 
properties. Disk-disk mergers tend to produce remnants with relatively
rapid rotation (Naab \& Burkert 2003; Naab, Jesseit \& Burkert 2006b),
whereas dry mergers tend to produce remnants with relatively little
rotation (Naab, Khochfar \& Burkert 2006a). A study of the evolution
of galaxy rotation properties can therefore contribute to an improved
understanding of the actual mechanisms that have shaped todays
elliptical galaxies.

The inferred rotation rates of the sample galaxies also provide the
potential to kinematically identify S0 galaxies that have been
visually misclassified as elliptical galaxies. S0 galaxies make up
only 12\% to 21\% of the galaxy population in the three clusters
studied here, compared to 50\% to 60\% for nearby clusters (Dressler
\etal 1997). In the three clusters, visually-classified elliptical
galaxies outnumber S0 galaxies by a ratio of $\sim 3:1$, whereas in
nearby clusters S0 galaxies outnumber elliptical galaxies. Issues of
misclassification have been raised as a potential contributing factor
to this strong apparent evolution of the S0 fraction (e.g., Fabricant
\etal 2000). This may be relevant for CL3C295 and CL0016+16, given
that their deficit of S0 galaxies is compensated by an overabundance
of elliptical galaxies (which constitute more than 50\% of the
population). In these two clusters the total early-type galaxy
fraction (i.e., elliptical and S0 galaxies combined) is only slightly
lower than in nearby clusters. By contrast, in CL1601+42 the majority
of galaxies are spirals, which outnumber the S0 galaxies by a factor
of 4--5.

The layout of the present paper is as follows. Section~\ref{s:models}
discusses the details of the modeling approach.
Section~\ref{s:results} presents the results for the individual
galaxies in the sample. Section~\ref{s:rot} discusses the inferred
rotation properties of the intermediate-redshift sample galaxies,
compares them to the rotation properties of early-type galaxies in the
local Universe, and discusses the results. Section~\ref{s:conc}
presents a summary and discussion of the paper's main conclusions.

The dynamical models also yield the mass-to-light ratios $M/L$ for all
the galaxies. These are used in van der Marel \& van Dokkum (2006,
hereafter Paper~II) to study the $M/L$ evolution of elliptical
galaxies as a function of redshift. This complements and tests
analyses based on Fundamental Plane evolution, such as those presented
in vDvdM06.

Throughout both papers we assume a cosmology with $\Omega_{\rm m} =
0.27$, $\Omega_{\Lambda} = 0.73$ (the values obtained by the Wilkinson
Microwave Anisotropy Probe; Spergel \etal 2003) and $H_0 = 71 \kms
\Mpc^{-1}$ (the value obtained by the HST Cepheid Key Project;
Section~7 of Freedman \etal 2001).

\section{Dynamical Models}
\label{s:models}

For the two-integral dynamical modeling we have used the assumptions
and software described previously by van der Marel \etal (1994b). The
goal is to first find a light density distribution that is consistent
with observed HST photometry, and then to find a dynamical solution of
the Jeans equations that reproduces the observed Keck spectroscopy.
Fitting the model to the kinematics yields both a normalized measure
$k$ of the galaxy's rotation rate and the dynamical $M/L$ over the
spectroscopically explored region.

The three-dimensional mass density is assumed to be oblate
axisymmetric and of the form
\begin{equation}
\label{rhodef}
  \rho(R,z) = \rho_0 (m/b)^{\alpha} [1+(m/b)^2]^{\delta} , \quad
                 m^2 \equiv R^2 + z^2 Q^{-2} .
\end{equation}
Here $z$ is the symmetry axis of the galaxy and $(R,\phi,z)$ are the
usual cylindrical coordinates. All isodensity surfaces are spheroids
with axial ratio $Q \leq 1$. The adopted mass density has no
particular physical significance. It is merely a convenient
parameterization in which the parameters $\rho_0$, $b$, $\alpha$ and
$\delta$ provide sufficient freedom to adequately fit the data
available for the sample galaxies. The components of the force ${\vec
\nabla} \Phi$ (where $\Phi$ is the gravitational potential) exerted by
the matter with mass density given in equation~(\ref{rhodef}) can be
expressed as one-dimensional integrals (Binney \& Tremaine 1987,
eq.~[2-88]) that are easily evaluated numerically.

The Jeans equations for hydrostatic equilibrium in a two-integral model
\begin{eqnarray}
\label{jeans}
   & & {{\partial \rho \overline{v_z^2}} \over {\partial z}} + 
          \rho \> {{\partial \Phi}\over{\partial z}} = 0 , \nonumber \\
   & & {{\partial \rho \overline{v_R^2}} \over {\partial R}} + 
          \rho \> {{\partial \Phi}\over{\partial R}} + 
    {{\rho}\over{R}} \Bigl [ \overline{v_R^2} - \overline{v_{\phi}^2} \Bigr ]
    = 0 ,
\end{eqnarray}
form a closed set for the two unknowns $\overline{v_{\phi}^2}$ and
$\overline{v_R^2} \equiv \overline{v_z^2}$. For known $\rho$ and
${\vec \nabla} \Phi$ the equations can be solved numerically on a
two-dimensional grid in the meridional plane. The amount of mean
streaming in the model is unconstrained by the condition of
hydrostatic equilibrium. We therefore introduce a free parameter $k$
that assigns an amount of rotational support according to
\begin{equation}
\label{satohk}
   \overline{v_{\phi}} = k [\overline{v_{\phi}^2} - \overline{v_R^2}]^{1/2} .
\end{equation}
For $k=0$ the model is non-rotating, whereas for $|k|=1$ the velocity
dispersion tensor is isotropic and the model is a so-called ``oblate
isotropic rotator''. The maximum value of $|k|$ for a physical model
is defined by the criterion that $\sigma_{\phi}^2 \equiv
\overline{v_{\phi}^2} - \overline{v_{\phi}}^2$ is everywhere positive.
This implies that
\begin{equation}
\label{satohkmax}
   |k| \leq k_{\rm max} = \min_{(R,z)} \>\> [\> \overline{v_{\phi}^2} / 
                          (\overline{v_{\phi}^2} - \overline{v_R^2}) \> ]^{1/2} .
\end{equation}

For line-of-sight projection we adopt a Cartesian coordinate system
$(x,y,w)$ with $w$ along the line of sight, $(x,y)$ in the plane of
the sky, and $x$ along the major axis. The projected mass density is
\begin{equation}
\label{Sigmadef}  
   \Sigma(x,y) = \int_{-\infty}^{\infty} \rho(x,y,w) \> dw .
\end{equation}
We assume that the mass-to-light ratio $M/L$ (expressed throughout
this paper in units of $\Msun/\Lsun$) is constant with radius in the
galaxy. This is known to be a reasonable assumption when modeling
kinematical data inside the effective radius of elliptical galaxies
(van der Marel 1991; Kronawitter \etal 2000). The projected light
density is then $\mu(x,y) = \Sigma(x,y) / (M/L)$. When viewed at
inclination angle $i$, the projected contours are similar ellipses
with projected axial ratio $Q'$, with
\begin{equation}
\label{Qprimedef}
  Q'^2 \equiv \cos^2 i + Q^2 \sin^2 i .
\end{equation}
Two models of the same Q' and different inclination predict exactly
the same projected light density if they have the same values of $b$,
$\alpha$, $\delta$, and $j_0 Q$, where $j_0 \equiv \rho_0/(M/L)$ is a
luminosity density.

The projected luminosity-weighted line-of-sight velocity moments of
order $i$ are given by
\begin{equation}
\label{projmom}
  \langle \overline{v_w^i} \rangle (x,y) = {1\over{\Sigma(x,y)}} 
    \int_{-\infty}^{\infty} \rho(x,y,w) \> \overline{v_w^i} (x,y,w) \> dw .
\end{equation}
In the present context we are interested only in the first two
moments. For a two-integral system 
\begin{eqnarray}
\label{intlosmom}
   \overline{v_w} & = & \overline{v_{\phi}} \cos \phi \sin i , \nonumber \\
   \overline{v_w^2} & = & (\overline{v_{\phi}^2} - \overline{v_R^2})
        \cos^2 \phi \sin^2 i + \overline{v_R^2} .
\end{eqnarray}
The resulting line-of-sight integrals can be evaluated numerically
once the Jeans equations have been solved.

The observational setup is characterized by the slit width $W$, the
pixel scale $P$ in the spatial direction, and the seeing FWHM $S$. The
model predictions $\langle \overline{v_w^i} \rangle(x,y)$ must
therefore be convolved (in an intensity-weighted sense) both with a
square tophat function of size $W \times P$ and with the seeing
point-spread function (PSF). The latter was assumed to be Gaussian.
For the convolutions we used a Monte-Carlo integration procedure. This
ultimately yields for each spatial pixel (or set of binned pixels)
along the slit the model predictions for the mean velocity $V$ and for
the second velocity moment $\sigma_{\rm RMS}^2 \equiv V^2 + \sigma^2$,
where $\sigma$ is the velocity dispersion.
  
For a given inclination, the model parameters $j_0$, $b$, $\alpha$,
$\delta$ and $Q$ can be fixed by fitting the projected light density
$\mu(x,y)$ to the available galaxy imaging, which was calibrated to
the rest-frame $B$-band. To transform the best-fitting luminosity
density of each model to units of $\Lsun
\pc^{-3}$ we adopted a solar absolute B-band magnitude $M_{B,\odot} =
5.48$ (Binney \& Merrifield 1998) and the relation between arcsec and
pc dictated by the adopted redshift and cosmology. For the model thus
obtained we solve the Jeans equations using the trial values $M/L = 1$
for the mass-to-light ratio and $k=1$ for the rotational support. The
results are projected along the line-of-sight and convolved with the
observational setup. Let the predictions thus obtained be ${\hat V}$
and ${\hat \sigma}_{\rm RMS}$. The predictions for other values of
$M/L$ and $k$ are then obtained by simple linear scaling: ${\hat V}
\rightarrow k \> (M/L)^{1/2} \> {\hat V}$ and ${\hat \sigma}_{\rm
RMS} \rightarrow (M/L)^{1/2} \> {\hat \sigma}_{\rm RMS}$. Let the
observed mean velocity and velocity dispersion be $V_{\rm obs}$ and
$\sigma_{\rm obs}$. The data-model comparison residuals are then
$\delta V = V_{\rm obs} - V_{\rm sys} - k \> (M/L)^{1/2} \> {\hat
V}$ and $\delta \sigma = \sigma_{\rm obs} - [(M/L) {\hat
\sigma}_{\rm RMS}^2 - k^2 \> (M/L) {\hat V}^2]^{1/2}$. Here $V_{\rm
sys}$ is the systemic velocity of the galaxy. We minimize these
residuals in a $\chi^2$ sense (i.e., squared, weighted with the formal
uncertainties $\Delta V_{\rm obs}$ and $\Delta \sigma_{\rm obs}$, and
summed over all data points) to obtain the best-fitting values of
$M/L$, $k$, and $V_{\rm sys}$. The formal uncertainties on these
quantities (i.e., those that reflect propagation of random
uncertainties) follow from standard $\chi^2$ theory. If the formal
best fit for $|k|$ exceeds the maximum value $k_{\rm max}$ allowed for
a physical model by equation~(\ref{satohkmax}), then we adopt the
latter. In discussions of $V_{\rm obs}$ in the remainder of this paper
it is to be understood that $V_{\rm sys}$ was always subtracted. The
systemic velocities are of little interest by themselves (in the
present context) and will not be discussed further.

%%% TABLE 1 %%%

\newcommand{\tablecontentsgalprop}{
\tablecaption{Intermediate-Redshift Galaxy Sample\label{t:galprop}} 
\tablehead{
\colhead{ID} & \colhead{Type} & \colhead{$z_{\rm mod}$} & 
\colhead{${\rm PA}_{\rm maj}$} & \colhead{$\epsilon$} & 
\colhead{$Q$} & \colhead{$i$} & \colhead{FWHM} &
\colhead{$\sigma_{\rm eff}$} & \colhead{$\Delta \sigma_{\rm eff}$} &
\colhead{$M/L$} & \colhead{$\Delta M/L$} &
\colhead{$k$} & \colhead{$\Delta_{-} k$} & \colhead{$\Delta_{+} k$} \\ 
\colhead{(1)} & \colhead{(2)} & \colhead{(3)} & \colhead{(4)} &
\colhead{(5)} & \colhead{(6)} & \colhead{(7)} & \colhead{(8)} &
\colhead{(9)} & \colhead{(10)} & \colhead{(11)} & \colhead{(12)} &
\colhead{(13)} & \colhead{(14)} & \colhead{(15)} \\
}
\startdata
CL 3C295-568  & S0/Sb & 0.456 & 355 & 0.58 & 0.42 & 86.1 & 0.75 &  139 &  9 & 3.56 & 0.29 & 1.08$^*$ & 0.01 & 0.00 \\
CL 3C295-834  & E     & 0.456 &   2 & 0.08 & 0.73 & 34.9 & 0.75 &  166 & 14 & 2.70 & 0.33 & 1.22     & 0.28 & 0.00 \\
CL 3C295-868  & S0    & 0.456 & 228 & 0.16 & 0.74 & 53.9 & 0.75 &  190 & 18 & 5.87 & 0.76 & 0.30$^*$ & 0.36 & 0.36 \\
CL 3C295-968  & E     & 0.456 &  55 & 0.08 & 0.73 & 36.2 & 0.75 &  198 & 12 & 7.09 & 0.56 & 1.05     & 0.29 & 0.29 \\
CL 3C295-2014 & E/S0  & 0.456 & 327 & 0.19 & 0.73 & 58.5 & 0.75 &  231 & 18 & 5.93 & 0.63 & 1.72     & 0.56 & 0.04 \\
CL 0016-438   & E     & 0.546 & 358 & 0.08 & 0.73 & 34.2 & 0.91 &  216 & 13 & 3.78 & 0.35 & 0.42     & 0.44 & 0.44 \\
CL 0016-461   & E     & 0.546 & 312 & 0.33 & 0.62 & 72.7 & 0.79 &  262 & 23 & 6.51 & 0.56 & 0.50$^*$ & 0.23 & 0.23 \\
CL 0016-531   & E     & 0.546 &  71 & 0.23 & 0.70 & 63.8 & 0.79 &  208 & 20 & 3.71 & 0.55 & 0.38$^*$ & 0.47 & 0.47 \\
CL 0016-611   & E     & 0.546 & 143 & 0.14 & 0.75 & 49.1 & 0.91 &  163 & 17 & 2.59 & 0.46 & 1.30     & 0.46 & 0.00 \\
CL 0016-612   & E     & 0.546 & 319 & 0.13 & 0.75 & 46.9 & 0.91 &  265 & 37 & 4.19 & 0.42 & 0.55     & 0.65 & 0.65 \\
CL 0016-650   & E/S0  & 0.546 &  47 & 0.10 & 0.74 & 41.3 & 0.79 &  139 & 16 & 4.45 & 0.71 & 0.35$^*$ & 0.45 & 0.45 \\
CL 0016-659   & E     & 0.546 &  90 & 0.22 & 0.71 & 62.0 & 0.79 &  249 & 26 & 5.74 & 0.82 & 0.08$^*$ & 0.25 & 0.25 \\
CL 0016-724   & E     & 0.546 & 200 & 0.06 & 0.73 & 30.8 & 0.91 &  210 & 14 & 6.77 & 0.64 & 0.08     & 0.35 & 0.35 \\
CL 0016-725   & E     & 0.546 & 260 & 0.17 & 0.74 & 56.3 & 0.91 &  178 & 14 & 5.38 & 0.57 & 0.63$^*$ & 0.20 & 0.20 \\
CL 0016-745   & E     & 0.546 & 174 & 0.12 & 0.75 & 45.6 & 0.91 &  114 & 13 & 1.89 & 0.29 & 1.38     & 0.30 & 0.00 \\
CL 0016-2050  & E     & 0.546 & 286 & 0.15 & 0.74 & 52.5 & 0.79 &  121 & 17 & 2.64 & 0.64 & 1.34$^*$ & 0.09 & 0.00 \\
CL 1601-270   & S0/E  & 0.510 & 164 & 0.33 & 0.62 & 72.8 & 0.88 &  176 & 11 & 3.26 & 0.27 & 0.38$^*$ & 0.16 & 0.16 \\
CL 1601-292   & E     & 0.539 &  55 & 0.19 & 0.73 & 58.1 & 0.71 &  169 & 11 & 2.77 & 0.23 & 1.37     & 0.21 & 0.00 \\
CL 1601-524   & E     & 0.539 &  11 & 0.13 & 0.75 & 47.1 & 0.71 &  210 & 15 & 5.65 & 0.42 & 0.83$^*$ & 0.31 & 0.31 \\
CL 1601-619   & E     & 0.539 & 255 & 0.25 & 0.68 & 65.6 & 0.88 &  211 & 11 & 5.66 & 0.42 & 1.04$^*$ & 0.24 & 0.24 \\
CL 1601-753   & E     & 0.539 &  75 & 0.16 & 0.74 & 54.5 & 0.71 &  248 & 11 & 6.19 & 0.25 & 0.03$^*$ & 0.12 & 0.12 \\
CL 1601-814   & E     & 0.539 & 230 & 0.28 & 0.67 & 68.2 & 0.71 &  219 & 19 & 4.98 & 0.47 & 1.12$^*$ & 0.27 & 0.11 \\
CL 1601-2040  & E     & 0.539 & 184 & 0.39 & 0.57 & 75.7 & 0.88 &  229 &  9 & 4.74 & 0.24 & 0.39$^*$ & 0.07 & 0.07 \\
CL 1601-2043  & E     & 0.539 &   2 & 0.21 & 0.71 & 61.3 & 0.71 &  158 & 14 & 3.45 & 0.52 & 0.49     & 0.39 & 0.39 \\
CL 1601-2060  & E     & 0.539 & 336 & 0.25 & 0.69 & 65.1 & 0.71 &  218 & 33 & 5.91 & 0.56 & 1.17$^*$ & 0.10 & 0.08 \\
\enddata}

\newcommand{\tablecommgalprop}{Column~(1) lists the galaxy 
ID and column~(2) the morphological type, both from Smail \etal
(1997). Column~(3) lists the redshift used for the modeling. For
cluster members this is the redshift of the cluster from Dressler \&
Gunn (1992) and Dressler \etal (1999); for the field galaxy CL
1601-270 it is the observed redshift reported in vDvdM06. Column~(4)
lists the major axis position angle and column~(5) the apparent
ellipticity ($\epsilon = 1 - Q'$), both determined as described in
Section~\ref{ss:phot}. Column~(6) lists the intrinsic axial ratio $Q$
of our ``standard'' model, as defined by
equation~(\ref{meanQ}). Column~(7) lists the corresponding
inclination, calculated from equation~(\ref{Qprimedef}). Column~(8)
lists the seeing FWHM of the spectroscopic observations in arcsec,
determined by modeling the observed galaxy intensity profiles along
the slit as described in Section~\ref{ss:kin}. For the four galaxies
that were observed with two different slit masks we list only the best
value of the seeing. Columns~(9) and~(10) list the velocity dispersion
for an aperture with the size of the effective radius, and its formal
uncertainty. Columns~(11)--(15) list the quantities inferred from our
dynamical models with the standard axial ratio and
inclination. Columns~(11) and~(12) list the best-fitting mass-to-light
ratio (in solar B-band units) and its formal
uncertainty. Columns~(13)-(15) list the best-fitting rotational
support and its formal uncertainties
$k_{-\Delta_{-}k}^{+\Delta_{+}k}$. If ${\Delta_{+}k} = 0$ then this
indicates that the best fit is obtained when $k$ is set to the maximum
value that guarantees positivity of $\sigma_{\phi}^2$, as given by
equation~(\ref{satohkmax}). An asterisk in column~(13) indicates that
the galaxy was included in the rotational support analysis presented
in Section~\ref{s:rot}.}

\ifemulate
\begin{deluxetable*}{llcrcccccccclcc}
\tabletypesize{\tiny}
\tablecontentsgalprop
\tablecomments{\small\tablecommgalprop}
\end{deluxetable*}
\fi

%%% END TABLE %%%

%%% FIGURE %%%

\newcommand{\figcapphot}{Elliptical isophote fit results for 
the sample galaxies as a function of semi-major axis length. Shown are
from left to right: isophotal surface brightness, transformed to
rest-frame B-band and corrected for $(1+z)^4$ cosmological surface
brightness dimming; isophotal ellipticity; and major-axis position
angle. The curves in the surface brightness panels show the
predictions of the best-fit mass density models of the form given by
equation~(\ref{rhodef}). The horizontal lines in the ellipticity and
position angle panels show the constant values used in the dynamical
modeling (these are the weighted average values determined as
described in Section~\ref{ss:phot}).\label{f:phot}}

\ifemulate

\begin{figure*}[t]
\epsfxsize=0.8\hsize
\centerline{\epsfbox{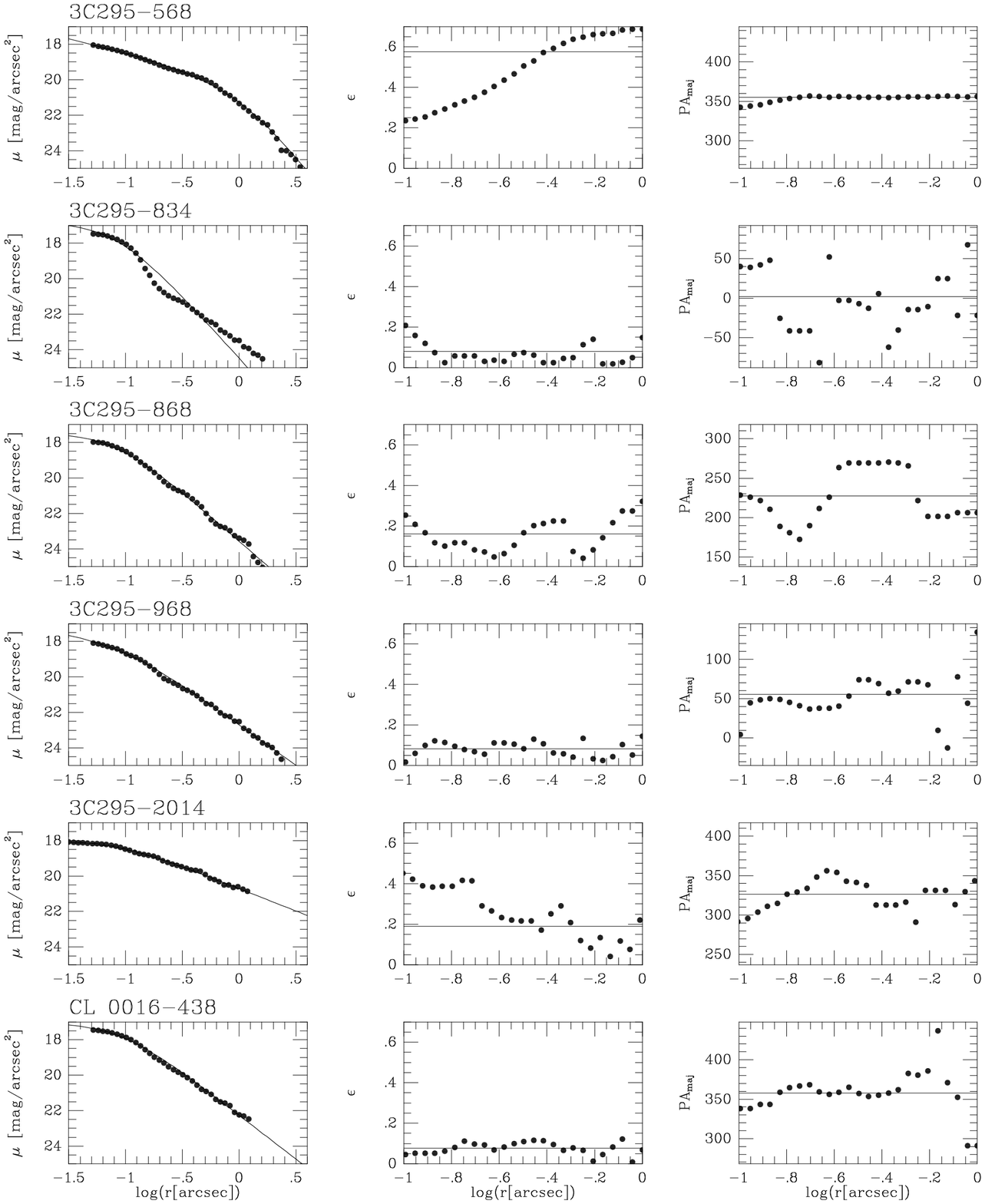}}
\figcaption{\figcapphot}
\addtocounter{figure}{-1}
\end{figure*}

\begin{figure*}[t]
\epsfxsize=0.8\hsize
\centerline{\epsfbox{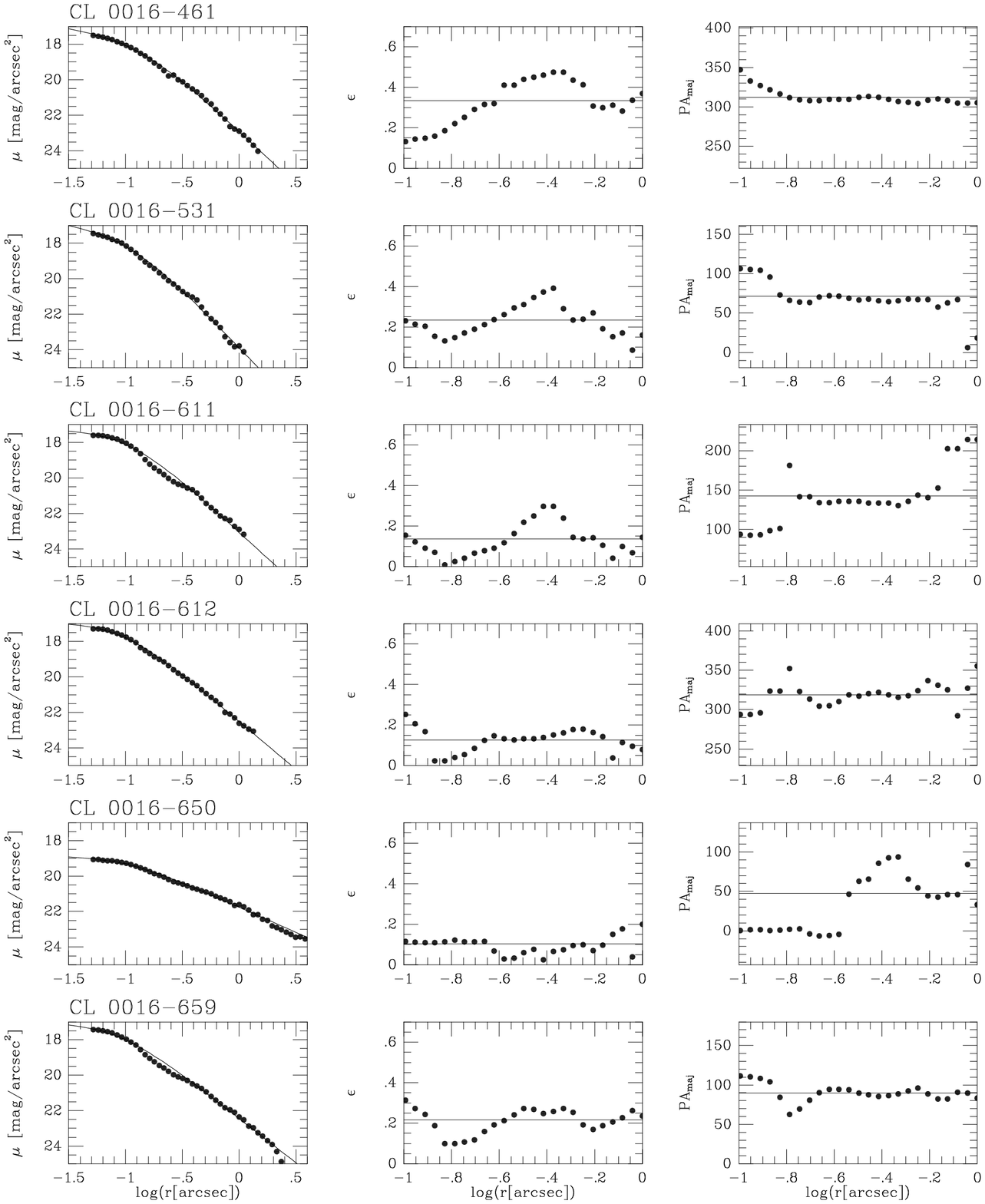}}
\figcaption{(continued)}
\addtocounter{figure}{-1}
\end{figure*}

\begin{figure*}[t]
\epsfxsize=0.8\hsize
\centerline{\epsfbox{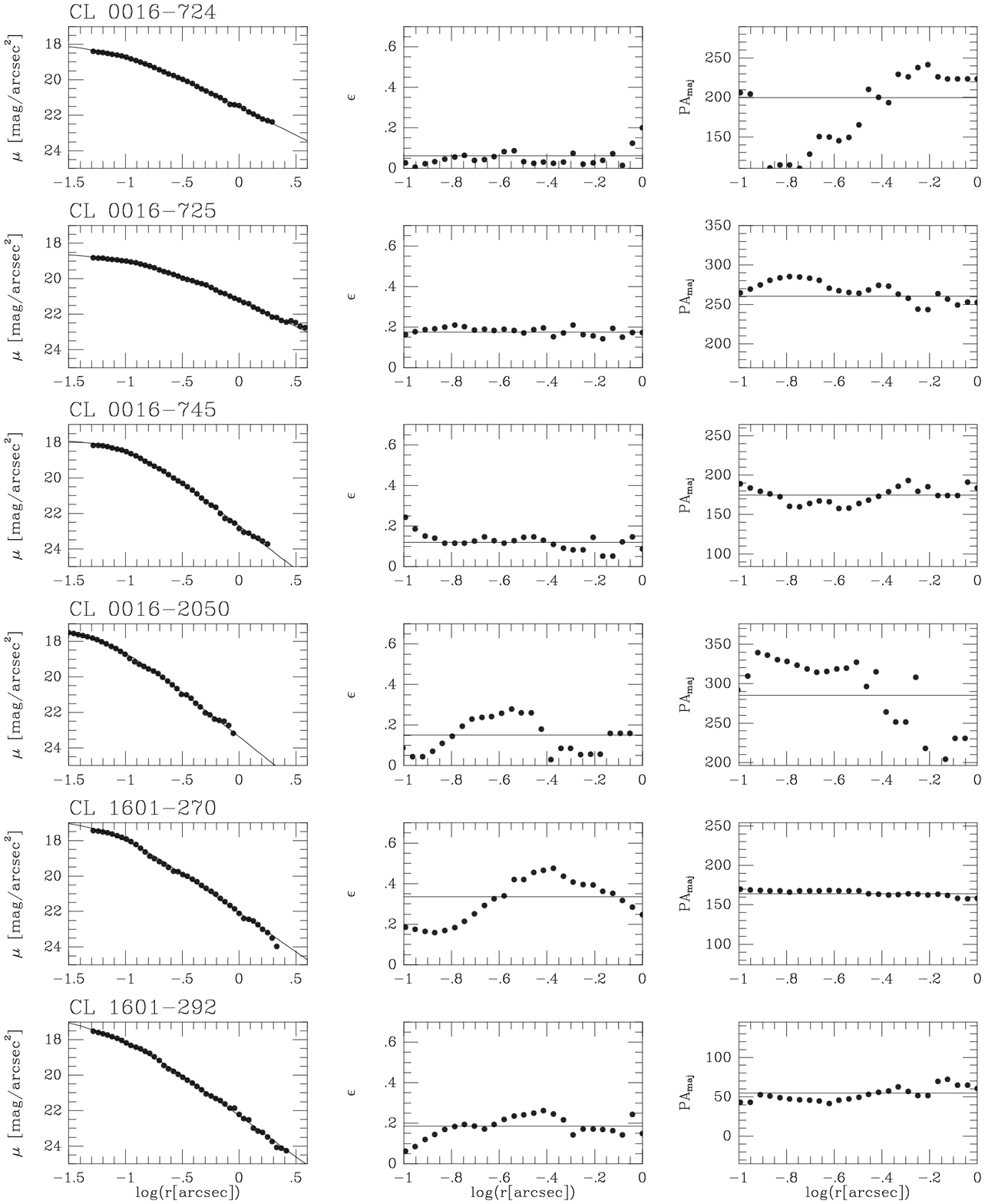}}
\figcaption{(continued)}
\addtocounter{figure}{-1}
\end{figure*}

\begin{figure*}[t]
\epsfxsize=0.8\hsize
\centerline{\epsfbox{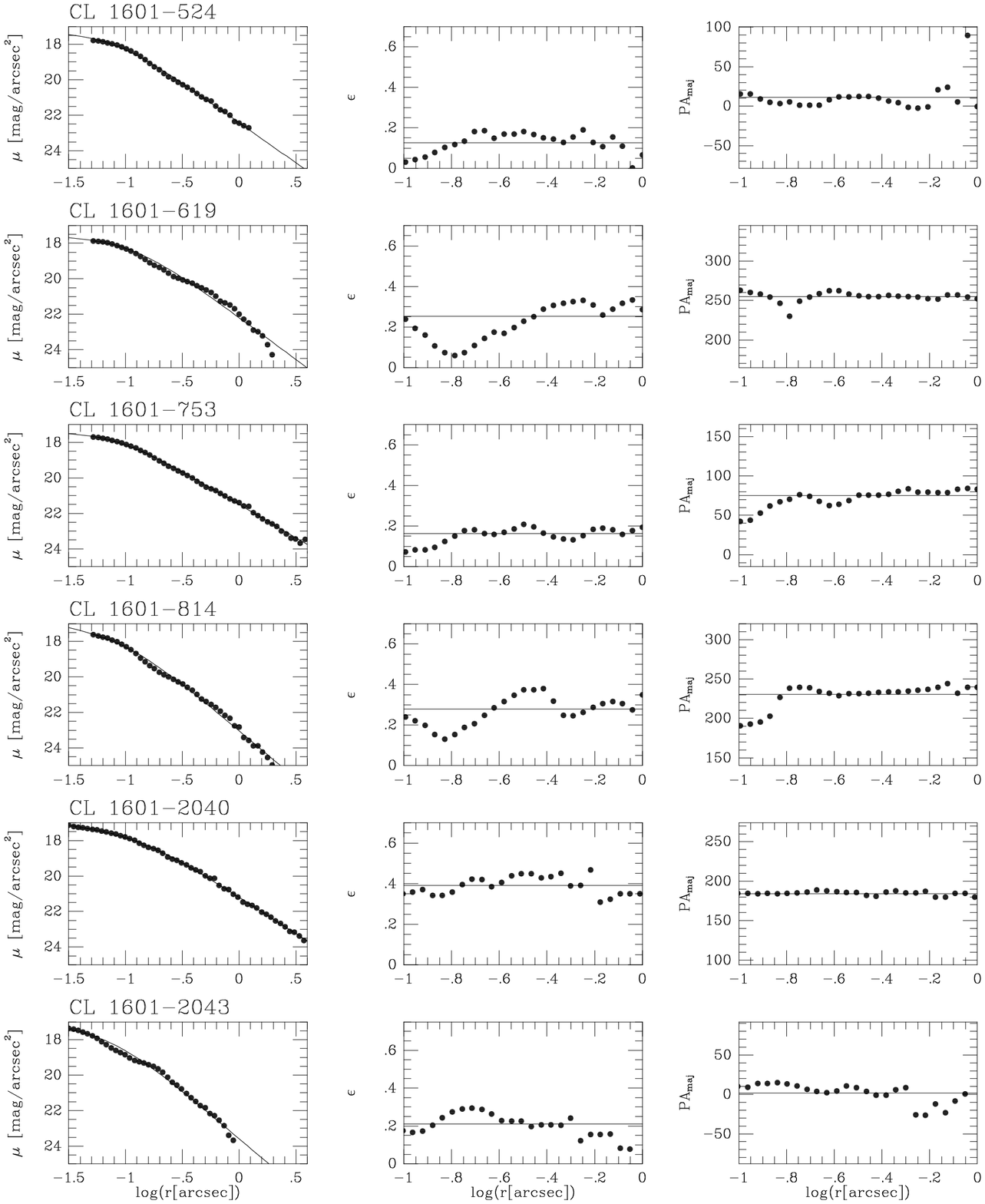}}
\medskip
\centerline{\epsfbox{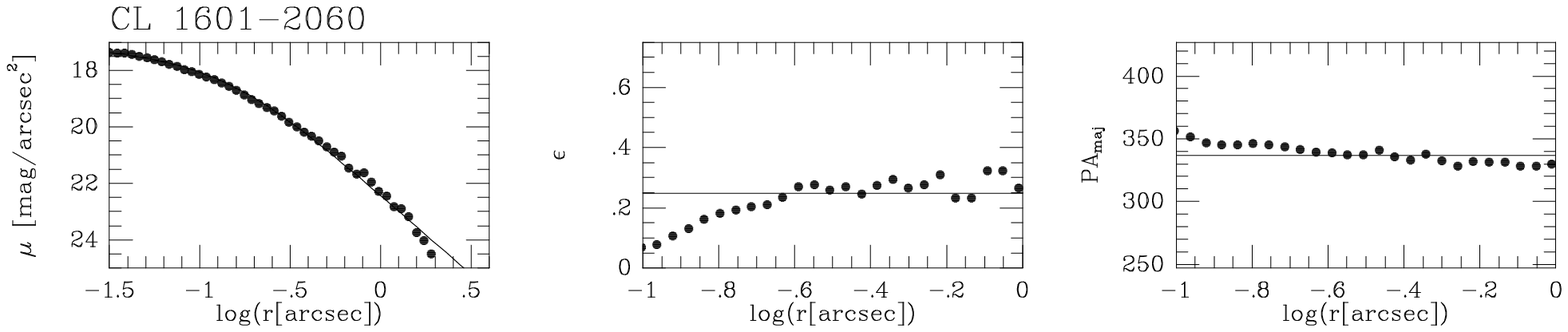}}
\figcaption{(continued)}
\end{figure*}

\fi

%%% END FIGURE %%%

%%% FIGURE %%%

\newcommand{\figcapkin}{Kinematics of the sample galaxies. Each set of two 
adjacent panels show the rotation velocity $V$ and the velocity
dispersion $\sigma$ along a given slit. The galaxy name is indicated
in the label above the rotation velocity panel. The slit position
angle is indicated in parentheses and is expressed as the sum of the
major axis position angle given in Table~\ref{t:galprop} and an offset
angle. The abscissa is given in arcsec. Data points are indicated with
error bars. The heavy solid curves show the predictions of the
best-fitting models with the standard inclination (see
Table~\ref{t:galprop}), taking into account the observational setup
and spatial resolution. For comparison, the dashed curves show the
model predictions as they would appear at infinite spatial resolution.
\label{f:kin}}

\ifemulate

\begin{figure*}[t]
\epsfxsize=0.8\hsize
\centerline{\epsfbox{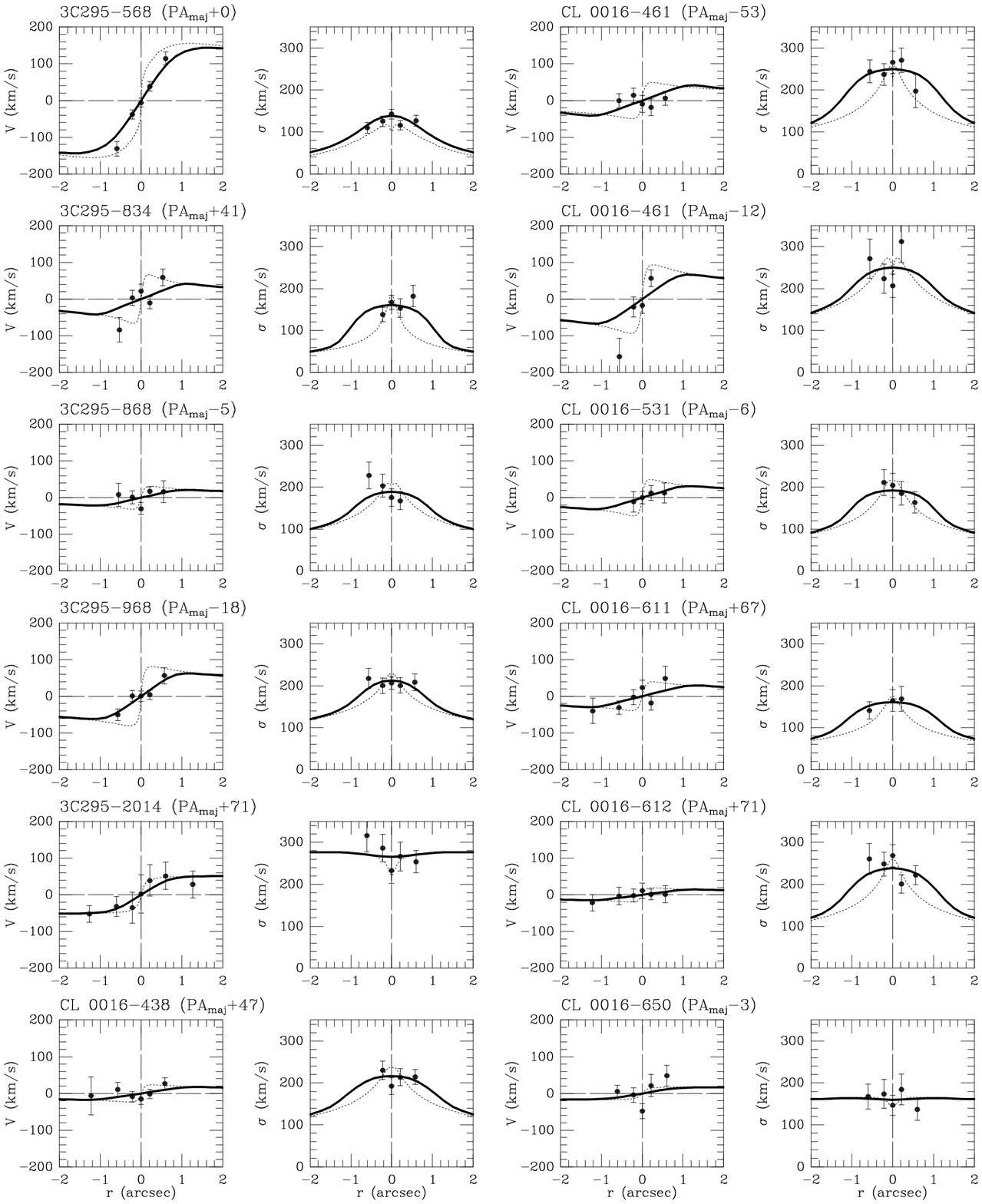}}
\figcaption{\figcapkin}
\addtocounter{figure}{-1}
\end{figure*}

\begin{figure*}[t]
\epsfxsize=0.8\hsize
\centerline{\epsfbox{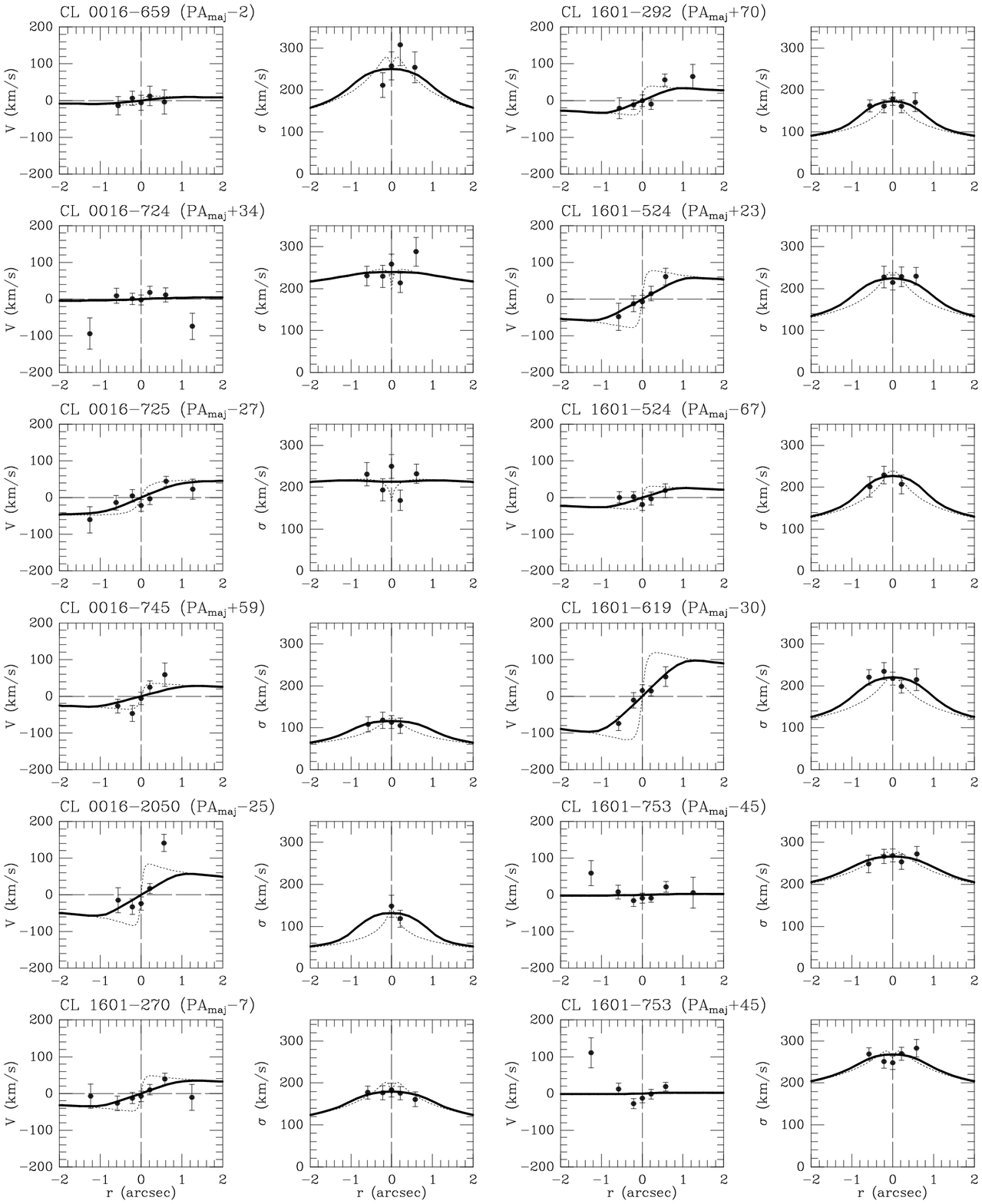}}
\figcaption{(continued)}
\addtocounter{figure}{-1}
\end{figure*}

\begin{figure}[t]
\epsfxsize=0.8\hsize
\centerline{\epsfbox{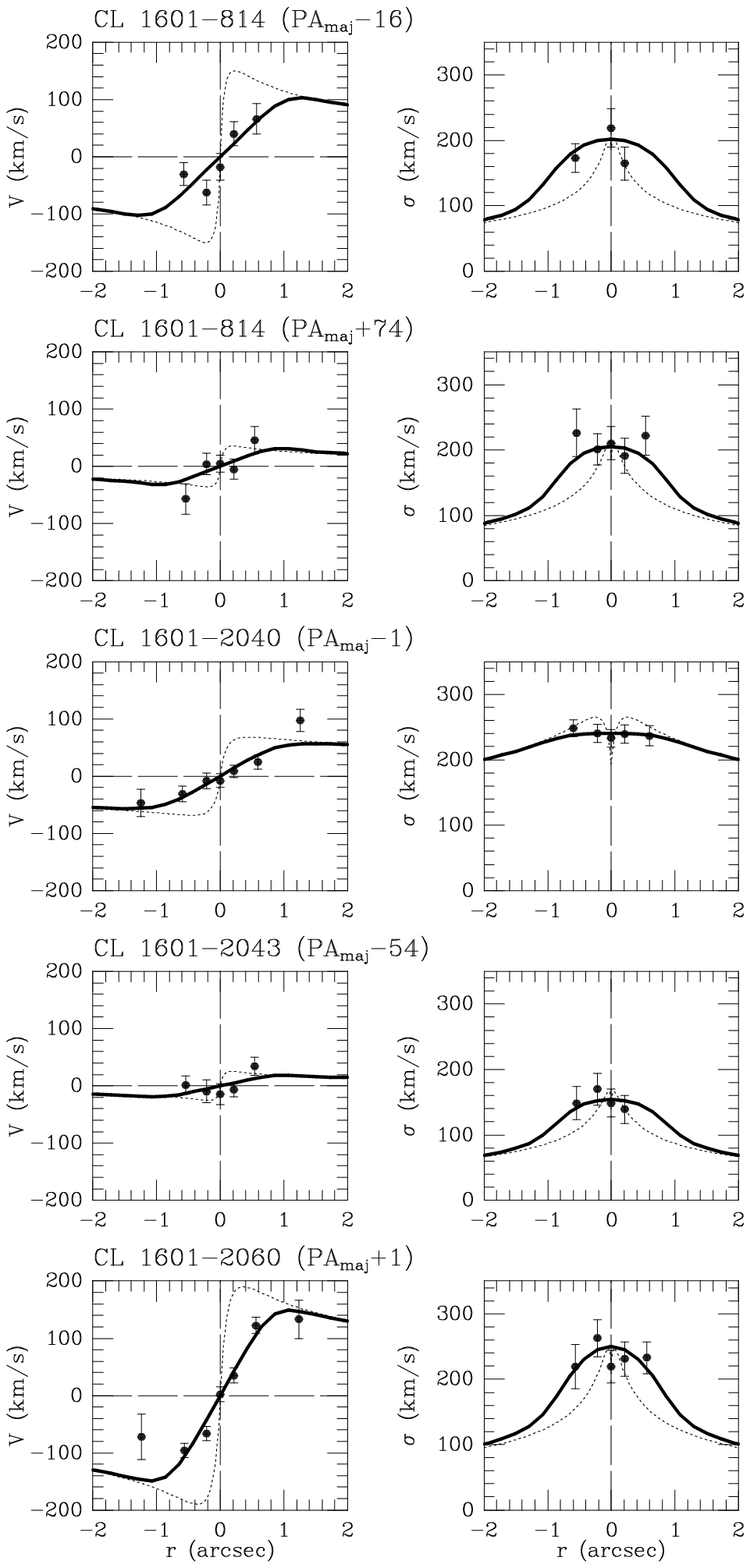}}
\figcaption{(continued)}
\end{figure}

\fi

%%% END FIGURE %%%

The only parameter that is not generally constrained by the data is
the inclination, which determines the axial ratio $Q$ according to
equation~(\ref{Qprimedef}). One possibility is to assume that all
galaxies are viewed edge-on, so that $i=90^{\circ}$ and the intrinsic
axial ratio is equal to the (observed) projected axial ratio:
$Q=Q'$. Alternatively, one can assume that all galaxies have a small
and fixed axial ratio, $Q=Q_{\rm min}$; we have taken $Q_{\rm min} =
0.4$ which is approximately the smallest intrinsic axial ratio found
for early-type galaxies (Tremblay \& Merritt 1995). The edge-on and
$Q=Q_{\rm min}$ models are useful in that they produce, respectively,
the roundest and flattest models possible, so that they bracket the
range of available possibilities. However, since they both represent
extremes, they are less useful for calculating the average properties
of a large sample. For that purpose it is better to use models that
are ``correct'' in an average statistical sense. To construct such
models we used available knowledge on the projected axial ratio
distribution $f(Q')$ of nearby early-type galaxies. Tremblay \&
Merritt (1995) inverted this distribution to obtain the distribution
of intrinsic axial ratios $f(Q)$. They did so under various
assumptions for the underlying geometry, and we use here their results
for an oblate axisymmetric geometry. Given $f(Q)$, one can
find\footnote{We calculated $f(Q | Q')$ with a simple Monte-Carlo
scheme in which for a very large number of hypothetical galaxies we
perform the following steps: (a) draw a random three-dimensional
viewing direction; (b) draw an intrinsic axial ratio $Q$ from the
distribution $f(Q)$; (c) calculate $Q'$ from
equation~(\ref{Qprimedef}); (d) add the results to a joint frequency
distribution of the quantities $Q$ and $Q'$.} the distribution $f(Q |
Q')$ that a galaxy of known projected axial ratio $Q'$ has intrinsic
axial ratio $Q$. In turn, this can be used to calculate the mean
intrinsic axial ratio as function of projected axial ratio $Q'$:
\begin{equation}
\label{meanQ}
  \langle Q \rangle (Q') \equiv \int Q f(Q | Q') \> dQ . 
\end{equation}
In our models we set the intrinsic axial ratio $Q$ of each galaxy to
the value $\langle Q \rangle (Q')$ appropriate for its observed
projected axial ratio $Q'$. We calculate the corresponding inclination
from equation~(\ref{Qprimedef}). We will refer to the values thus
obtained as the ``standard'' axial ratio and inclination. Results
discussed in the remainder of this paper are for these standard
values. Our approach does not guarantee that we use the correct
inclination for any individual galaxy. However, it does imply that our
models have the correct average intrinsic axial ratios when averaged
over a large sample. Where relevant, we address the dependence of the
results on inclination by referring to results obtained with the
edge-on models or $Q=Q_{\rm min}$ models.

\section{Results for Individual Galaxies}
\label{s:results}

\subsection{Sample}
\label{ss:sample}

For the modeling we used the sample of galaxies from vDvdM06. The
galaxies reside in the clusters CL 3C295, CL 0016+1609, and CL
1601+4253, at redshifts $z = 0.456$, $0.546$, and $0.539$,
respectively (Dressler \& Gunn 1992; Dressler \etal 1999). The
clusters were selected based on their visibility at the time of the
Keck observations, and because they are among the most S0 deficient
clusters in the MORPHS sample (Dressler \etal 1997). The MORPHS sample
itself was not selected according to strict criteria. The galaxy
selection was largely constrained by the geometry of the Keck/LRIS
masks, and by the fact that sample galaxies should be bright enough
for spectroscopy. Priority was given to galaxies classified from HST
images as E or E/S0 by Smail \etal (1999). The latest-type galaxy
included in the sample was an S0/Sb galaxy. This galaxy 3C295-568 was
included for the specific purpose to see if rotation could be measured
reliably. For the present paper we excluded two galaxies from the
sample of vDvdM06: 3C295-47 and CL1601-474. The former is a foreground
field galaxy at $z=0.131$. The latter has a disturbed morphology
suggesting an ongoing merger, so that equilibrium models are likely to
be of limited value. One other galaxy in the sample of vDvdM06 was
found to be a field galaxy, namely CL 1601-270. This galaxy was
retained because its redshift $z=0.510$ is close to that of the other
galaxies.

The galaxies in the final sample are listed in Table~\ref{t:galprop},
including redshifts, visual classifications, and $\sigma_{\rm
eff}$. The effective velocity dispersions are included for use in
Paper~II. They were calculated from the observed velocity dispersions
measured from spectra that summed the central five pixels along each
slit (presented in vDvdM). These measurements were transformed to an
estimate of the dispersion inside an aperture of size $r_{\rm eff}$
using the formulae given in Jorgensen \etal (1995b). The values of
$r_{\rm eff}$ and $I_{\rm eff}$ for the sample galaxies can be found
in vDvdM06.
  
\subsection{Photometry}
\label{ss:phot}

To start the analysis we used the HST/WFPC2 images previously shown
and discussed in vDvdM06. For the galaxies in the clusters CL3C295 and
CL1601+42 there are only images in the F702W filter. For the galaxies
in the cluster CL0016+16 there are images in both F555W and F814W, but
we only used the latter. The images were corrected for the blurring of
the HST PSF using Lucy-Richardson deconvolution (Richardson 1972; Lucy
1974). The PSF used for the deconvolution was calculated with the
TinyTim software package (Krist \& Hook 2001). The deconvolved galaxy
images were analyzed through isophotal ellipse fitting with the task
{\tt ELLIPSE} in the {\tt IRAF} software package (Jedrzejewski
1987). This yields for each galaxy the major axis profile of surface
brightness, ellipticity, and position angle. The results are shown in
Figure~\ref{f:phot}. The profiles can generally be followed out to
1--5 arcsec from the galaxy center. The ellipticity and position angle
are not shown in the figure inside $0.1$ arcsec, where they are
generally unreliable due to the residual effects of the HST PSF. They
are also not shown outside of $1.0$ arcsec, where they tend to become
unreliable due to the low galaxy brightness relative to the sky
brightness.

The models described in Section~\ref{s:models} are axisymmetric with
constant axial ratio, and in projection have constant ellipticity
$\epsilon = 1 - Q'$ and major axis position angle ${\rm PA}_{\rm
maj}$. To optimally choose these parameters for our models we
calculated for each galaxy the average ellipticity and major axis
position angle between $0.1$ and $1$ arcsec, weighted both with the
flux and the annular area that each isophote represents. The results
are listed in Table~\ref{t:galprop} and are also shown as horizontal
lines in Figure~\ref{f:phot}. Clearly, the assumption of constant
ellipticity and position angle with radius is better for some galaxies
(e.g., CL1601-2040) than for others (e.g., CL0016-2050). However, even
for galaxies with significant ellipticity variations or isophotal
twists our approach still has the advantage that it properly takes
into account some average measure of galaxy flattening.

The observed surface brightnesses in counts per second per pixel were
transformed to magnitudes per square arcsecond using the known WFPC2
pixel sizes and filter zero-points from the WFPC2 Data Handbook
(Baggett \etal 2002). The results were corrected for foreground
Galactic extinction using the Schlegel \etal (1998) maps. The
magnitudes in the filters used for the observations were then
transformed to galaxy rest-frame B-band magnitudes using the relations
given in vDvdM06. And finally, a $(1+z)^4$ correction for cosmological
surface brightness dimming was applied.

For each galaxy we determined the parameters so that the projected
luminosity density $\mu(x,y)$ of the model defined by
equation~(\ref{rhodef}) best fits the observed surface brightness
profile. The fit was done using a Levenberg-Marquardt iteration scheme
(Press \etal 1992). Data points within 2 arcsec from the galaxy center
were included in the fit. The integrated brightness of the model in
the central pixel was also included. The predictions of the best model
fits are overplotted on the data in Figure~\ref{f:phot}. The
predictions generally fit the observed surface brightness profiles
reasonably well; the median residual over the entire sample is 0.09
mag/arcsec$^2$. The worst fit is obtained for CL 3C295-834, which
clearly has more structure in its profile than can be accounted for by
our model. The adopted inclination $i$ and axial ratio $Q$ for our
standard model of each galaxy are listed in Table~\ref{t:galprop}. For
a given galaxy, the luminosity density $j_0$ is inversely proportional
to $Q$ because the total luminosity of the model, which is fixed by
the observations, is proportional to $j_0 Q$. However, the inferred
parameters $b$, $\alpha$, and $\delta$ (see eq.~[\ref{rhodef}]) and
the fits shown in Figure~\ref{f:phot} are all independent of the
assumed $i$ and $Q$.

The galaxy CL 3C295-2014 is the well-known AGN 3C295. There is the
possibility that this galaxy contains a central non-thermal point
source that could bias the analysis. However, the results for both the
light profile of this galaxy and its $M/L$ (see Figure~\ref{f:phot}
and Paper~II) do not provide any evidence for deviations from the
trends defined by the other galaxies in the sample. We therefore
retained CL 3C295-2014 in our sample and did not treat it in any
special way.

%%% FIGURE %%%

\newcommand{\figcapklum}{(a) $(v/\sigma)^*$ versus $B$-band absolute 
magnitude $M_B$ for nearby elliptical galaxies. The data are from the
compilations of Davies \etal (1983) and Bender, Burstein \& Faber
(1992). The $M_B$ values from these papers were transformed to $H_0 =
71 \kms \Mpc^{-1}$, for consistency with the value assumed here. (b)
$k$ versus $M_B$ for the sample of intermediate redshift cluster
galaxies. The values of $M_B$ were corrected to $z=0$ using the
luminosity evolution calculated by vDvdM06, for comparison with the
local sample. Symbol types indicate the visual galaxy classification
from Smail \etal (1997): E (solid dot), E/S0 (hexagon), S0/E (square),
S0 (triangle), or S0/Sa (open circle).  Galaxies with apparent
ellipticity $\epsilon < 0.10$ were excluded from both panels. For the
intermediate redshift cluster sample, galaxies were included only if
there is data along a slit within 45$^{\circ}$ from the major axis. In
both panels, galaxies with $M_B \lta -20.4$ span a range of
$(v/\sigma)^*$ or $k$ values. However, most of these high-luminosity
galaxies do not rotate fast enough to account for their flattening
(i.e., they have $(v/\sigma)^* < 1$ or $k < 1$).\label{f:klum}}

\ifemulate
\begin{figure*}[t]
\epsfxsize=0.8\hsize
\centerline{\epsfbox{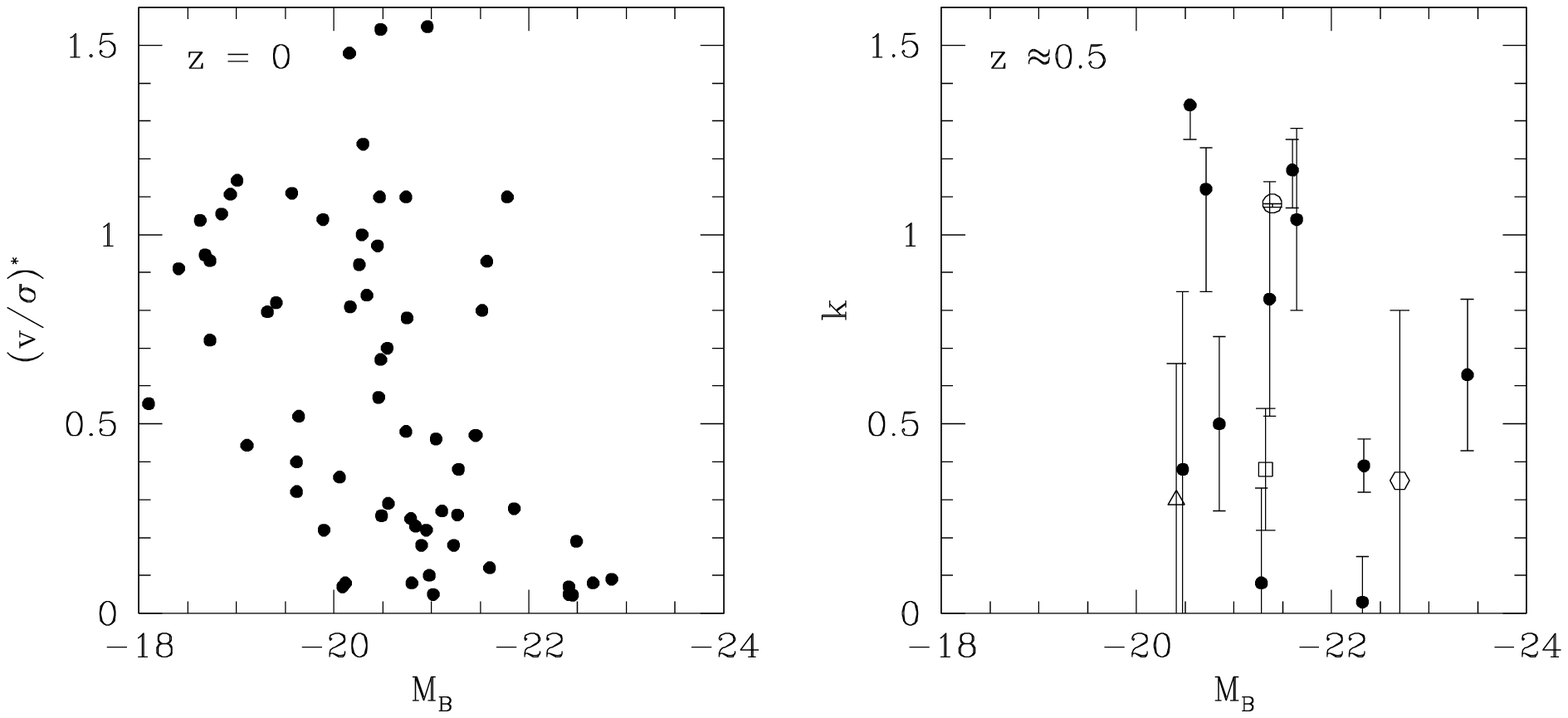}}
\figcaption{\figcapklum}
\end{figure*}
\fi

%%% END FIGURE %%%

\subsection{Kinematics}
\label{ss:kin}

For each galaxy we solved the Jeans equations to obtain predictions
for the projected kinematics on the plane of the sky. To model the
observational setup we used the known slit width $W = 1.1$ arcsec and
pixel scale $P = 0.215$ arcsec/pixel. Binning along the slit that
was applied to the data (see vDvdM06) was also taken into account. We
assume that during the observations all slits were perfectly centered
on their corresponding galaxy. This was probably true to within $\sim
0.2''$ accuracy. To determine the seeing for the observations with a
given slit mask we ran models with different values of the FWHM
$S$. We compared the observed and predicted intensity profiles along
the slit for the galaxies observed with that mask, and adopted the
value that gave the best agreement in a $\chi^2$ sense. The inferred
seeing FWHM values range from $0.71''$ to $0.91''$ (consistent with
estimates obtained from inspection of stars observed during the same
run) and are listed in Table~\ref{t:galprop}. The position angle of
each slit with respect to the galaxy major axis was determined by
subtracting the value ${\rm PA}_{\rm maj}$ in Table~\ref{t:galprop}
from the known slit position angle on the sky. The values of the
mass-to-light ratio $M/L$ and rotational support $k$ were determined
to optimize the fit to the observed kinematics along the slit. For
four galaxies (namely: CL0016-461, CL1601-524, CL1601-753, CL1601-814)
data were available for more than one slit position angle (from
observations with different slit masks). In those cases a single
combination of $M/L$ and $k$ was fit simultaneously to the data for
both slits.

Figure~\ref{f:kin} shows the resolved kinematical profiles for all
galaxies (obtained as described in Section 2.6 of vDvdM06), together
with the best fits of the standard-inclination models. The fits are
generally good, as confirmed by the $\chi^2$ values. When summed over
the entire sample, the mean velocity data points are fit with $\chi^2 =
145.8$ for 113 degrees of freedom. The velocity dispersion data points
are fit with $\chi^2 = 81.1$ for 102 degrees of freedom. Models
constructed with other assumptions for the inclination are both
visually and statistically indistinguishable from those constructed
for the standard inclinations. Comparison of edge-on models and models
with $Q=Q_{\rm min}$ yields a difference in total $\chi^2$, summed
over the entire sample, of only $\Delta \chi^2 = 1.0$. This indicates
that our kinematical data provide insufficient information to
constrain the galaxy inclinations. Indeed, that would require high
quality two-dimensional velocity fields (Cappellari \etal 2006).

For comparison, Figure~\ref{f:kin} also shows the model predictions
before convolution with the slit width, pixel size, and PSF. It is
clear that this convolution has a major impact on the predictions, due
to the fact that the spatial resolution of the observations is not
much smaller than the galaxies themselves. In particular: (a) the
observed rotation curves are always much shallower than the rotation
curves that would be observed with infinite spatial resolution; and
(b) the velocity dispersion profiles fall off less steeply with radius
than they would when observed with infinite spatial resolution. Both
effects are easily understood as the consequence of scattering of
light from small radii (where the galaxy is brightest) to larger
radii. Obviously, the resolved kinematics of distant galaxies can only
be meaningfully interpreted if detailed dynamical models are combined
with a proper accounting of the spatial resolution of the
observations, as we have done here.

The kinematical data typically extend to between $0.2''$ and $1.2''$
from the galaxy center with a typical value of $\sim 0.7''$. At $z =
0.5$ this corresponds to $4.3 \kpc$. So our data extend about as far
out as the data available for many nearby elliptical galaxies (e.g.,
van der Marel 1991; Emsellem \etal 2004). On the other hand, the
sampling and resolution of the kinematical profiles in physical units
are of course much worse. Nonetheless, for the mean velocities $V$ it
is clear that our spatially resolved measurements do provide important
new insight on the internal structure of the galaxies as compared to a
single integrated spectrum. The results for the S0/Sb galaxy 3C295-568
show that rotation can be reliable measured. While this is the most
rapidly rotating galaxy in the sample (as expected, given its visual
classification), most of the other galaxies also have statistically
significant gradients in $V$ along the slit. With the help of our
models this provides constraints on the intrinsic rotation
velocity. By contrast, for the velocity dispersions $\sigma$ it is not
clear that we are able to learn much more about the internal structure
than we would from a single integrated measurement. The model
predictions in Figure~\ref{f:kin} show that any gradients in $\sigma$
over the central arcsec are expected to be small. To discriminate
between different models one would need to be able to reliably measure
these small gradients. In most galaxies our formal uncertainties on
$\sigma$ are not small enough to allow this. Of course, we do measure
an accurate mass-to-light ratio, but that would have been possible
also with only a single integrated $\sigma$ measurement.

Table~\ref{t:galprop} lists the best-fitting values of $k$ and $M/L$
(in rest-frame $B$-band solar units) for the standard-inclination
models, with their formal uncertainties. The sign of $k$ is of little
interest and indicates merely in which of two possible directions the
galaxy is rotating around its symmetry axis. Adding $180^{\circ}$ to
the value of ${\rm PA}_{\rm maj}$ yields a reversal of the sign of
$k$. For convenience we have defined the values of ${\rm PA}_{\rm
maj}$ in Table~\ref{t:galprop} to be such that the best-fitting value
of $k$ is always positive. The $M/L$ values of the galaxies are not
addressed further here, since they are discussed and interpreted in
detail in Paper~II. Note that $M/L$ depends on the assumed distance,
and therefore on the assumed cosmological parameters $\Omega_{\rm m}$,
$\Omega_{\Lambda}$ and $H_0$ (as listed in Section~\ref{s:intro}). The
uncertainties in $M/L$ introduced by distance uncertainties are not
included in Table~\ref{t:galprop}, but they are quantified in
Paper~II.

\section{Galaxy Rotation Properties}
\label{s:rot}

\subsection{Intermediate-Redshift Cluster Sample}
\label{ss:intsample}

The model quantity $k$ is a normalized measure of the rotation rate of
a galaxy, with an oblate isotropic rotator model corresponding to
$k=1$. Inspection of Table~\ref{t:galprop} shows that the random
uncertainties in $k$ tend to be large when it is estimated from data
along a slit that is far from the major axis, or for any galaxy that
is nearly circular in projection on the sky. In these cases the value
of $k$ is also very sensitive to small systematic uncertainties in the
measured major axis position angle or the ellipticity of the
galaxy. For this reason we consider in the following only the 15
galaxies in the sample of intermediate redshift cluster galaxies for
which a slit was placed within $45^{\circ}$ of the major axis, and
which have projected ellipticity $\epsilon > 0.10$. These galaxies are
marked with an asterisk in column~(13) of Table~\ref{t:galprop}.

Figure~\ref{f:klum}b shows the inferred values of $k$ as a function of
the $B$-band absolute magnitude $M_B$. The latter was obtained from
the total $B$-band luminosity, which was estimated as $L_B = 2 \pi
r_{\rm eff}^2 I_{\rm eff}$. For comparison to local samples, the
luminosity was corrected to $z=0$ using the luminosity evolution
calculated by vDvdM06. The galaxies span a range of $k$ values between
$0$ to about $1.3$. The galaxies that were not visually classified as
$E$ galaxies are indicated with special symbols in the figure. The
galaxies CL 0016-650 (E/S0), CL 1601-270 (S0/E) and CL 3C295-868 (S0)
all have only modest rotation rates of $k=0.3$--$0.4$, albeit with
sizable uncertainties of similar magnitude. Either way, the rotation
rates of these galaxies do not make them stand out among the
ellipticals. By contrast, the galaxy CL 3C 295-568 (S0/Sb) has
well-measured ($\Delta k \leq 0.01$) rapid rotation ($k \approx
1.08$). This galaxy also distinguishes itself among the sample
galaxies by having the highest apparent flattening ($\epsilon =
0.58$).

\subsection{Comparison to Local Elliptical Galaxies}
\label{ss:localcomp}

The quantity that is most often measured for local galaxies to
quantify rotational support is $(v/\sigma)^{*}$. This is defined as
the ratio of the the observed $(v/\sigma)$ and the value of
$(v/\sigma)$ expected for an oblate isotropic rotator model with the
observed apparent flattening. This is almost identical to the
model parameter $k$ used here. The main difference is that
$(v/\sigma)^{*}$ is defined in terms of projected global quantities
(maximum rotation velocity and average velocity dispersion), whereas
$k$ is defined locally by equation~(\ref{satohk}). In principle this
difference in definition could lead to small systematic differences
between $k$ and $(v/\sigma)^{*}$, even when measured on the same
galaxy. However, these differences are unlikely to exceed $\sim
0.1$. This is negligible compared to the typical uncertainties in our
$k$ measurements for the intermediate redshift cluster galaxies (see
Figure~\ref{f:klum}b). In the following we therefore compare the
quantity $k$ directly to the quantity $(v/\sigma)^{*}$ available for
local galaxies.

As local comparison sample we combined the literature compilations of
$(v/\sigma)^{*}$ values for elliptical galaxies from Davies \etal
(1983) and Bender, Burstein \& Faber (1992). The resulting sample
contains 80 galaxies. For consistency with our treatment of the sample
of intermediate redshift cluster galaxies we excluded from the
following discussion the 12 galaxies with $\epsilon < 0.10$.

Figure~\ref{f:klum}a shows $(v/\sigma)^{*}$ for the local ellipticals
as a function of $M_B$. The local galaxies show a well-known trend
between $(v/\sigma)^{*}$ and $M_B$. Low-luminosity ellipticals ($M_B
\gta -20.0$) have values of $(v/\sigma)^{*}$ that cluster near
unity. This indicates that these galaxies are typically supported by
rotation. High-luminosity ellipticals ($M_B \lta -20.0$) have values
of $(v/\sigma)^{*}$ that range from zero to just above unity, but with
most of the galaxies clustering near zero. This indicates that these
galaxies are typically supported by velocity dispersion anisotropy
(for a more detailed discussion of what $(v/\sigma)^{*}$ measurements
imply for the velocity dispersion anisotropy of elliptical galaxies
see Cappellari \etal 2005). The three galaxies in
Figure~\ref{f:klum}a with $k \approx 1.5$ and $M_B$ between $-20.0$
and $-21.0$ are NGC 584, 2974, and 3640. These are probably not normal
ellipticals. NGC 584 is actually classified in the RSA catalog
(Sandage \& Tammann 1981) as an S0 galaxy, although it is an E4 in the
RC3. NGC 2974 is also an E4 galaxy, but it has all the hallmarks of
being a misclassified S0 galaxy (Cinzano \& van der Marel 1994). And
NGC 3640 has considerable fine-structure that suggests it is a recent
merger remnant (Prugniel \etal 1988).

The galaxies in the sample of intermediate redshift cluster galaxies
are all brighter than $M_B \approx -20.0$. The fact that they span a
wide range of $k$ values is therefore consistent with the behavior
seen in the local sample. Moreover, the four brightest galaxies ($M_B
< -22.0$) all appear to have insufficient rotation to account for
their flattening, consistent with the situation for the elliptical
galaxies in the nearby universe. However, the details of the
distribution of rotational support are somewhat different between the
local and distant samples. Figure~\ref{f:khist} compares a histogram
of the $k$ values for the intermediate redshift cluster galaxies to a
histogram of the $(v/\sigma)^{*}$ values of local galaxies. For the
intermediate redshift cluster galaxies we included only the 11
galaxies classified as ellipticals, for consistency with the local
sample. For the local galaxies we included only galaxies with $M_B <
-20.4$. With this selection, both histograms probe similar absolute
magnitude ranges (for the local sample $M_B$ has an average of $-21.2$
with a dispersion of $0.7$ and for the distant sample it has an
average of $-21.5$ with a dispersion of $0.9$).

The histograms in Figure~\ref{f:khist} show that in a relative sense
there is a higher number of rapidly rotating galaxies in the distant
sample than in the local sample. The average $\langle k
\rangle$ for the distant galaxies is $0.68 \pm 0.13$, whereas
the average $\langle (v/\sigma)^{*} \rangle$ for the local galaxies is
$0.47 \pm 0.07$. In the distant sample 36\% of the galaxies have $k >
0.9$, whereas in the local sample only 19\% have $(v/\sigma)^{*} >
0.9$.  This estimate of the percentage difference at high rotation
rates is actually conservative for two reasons. First, the percentage
of the local sample with $(v/\sigma)^{*} > 0.9$ would drop by 5\% if
we excluded the two potential S0 galaxies NGC 584 and 2974. And
second, values of $(v/\sigma)^{*}$ are defined as the rotational
support for an edge-on model. In contrast, our standard two-integral
models are somewhat inclined. If instead we had used the values of $k$
inferred from edge-on models for the intermediate redshift clusters,
then all $k$ values would have gone up by $\sim 7$\%. The discrepancy
between the histograms in Figure~\ref{f:khist} would then have been
even larger. In view of the statistics and these additional
considerations, we conclude that the difference between the rotation
properties of the galaxies in our $z \approx 0.5$ cluster sample and
those of our local comparison sample are significant. In the following
subsections we discuss possible explanations for this difference.

%%% FIGURE %%%

\newcommand{\figcapkhist}{Comparison of a histogram of the $k$
values for the intermediate redshift cluster galaxies (heavy line) to
a histogram of the $(v/\sigma)^{*}$ values of local galaxies (dotted
line). For the intermediate redshift cluster galaxies we included only
the 11 galaxies with reliable $k$ determinations that were classified
as ellipticals, for consistency with the local sample. For the local
galaxies we included only galaxies with $M_B < -20.4$, so that both
histograms probe similar absolute magnitude ranges. The histograms
show that in a relative sense there is a higher number of rapidly
rotating galaxies in the distant sample than in the local
sample.\label{f:khist}}

\ifemulate
\begin{figure}[t]
\epsfxsize=0.9\hsize
\centerline{\epsfbox{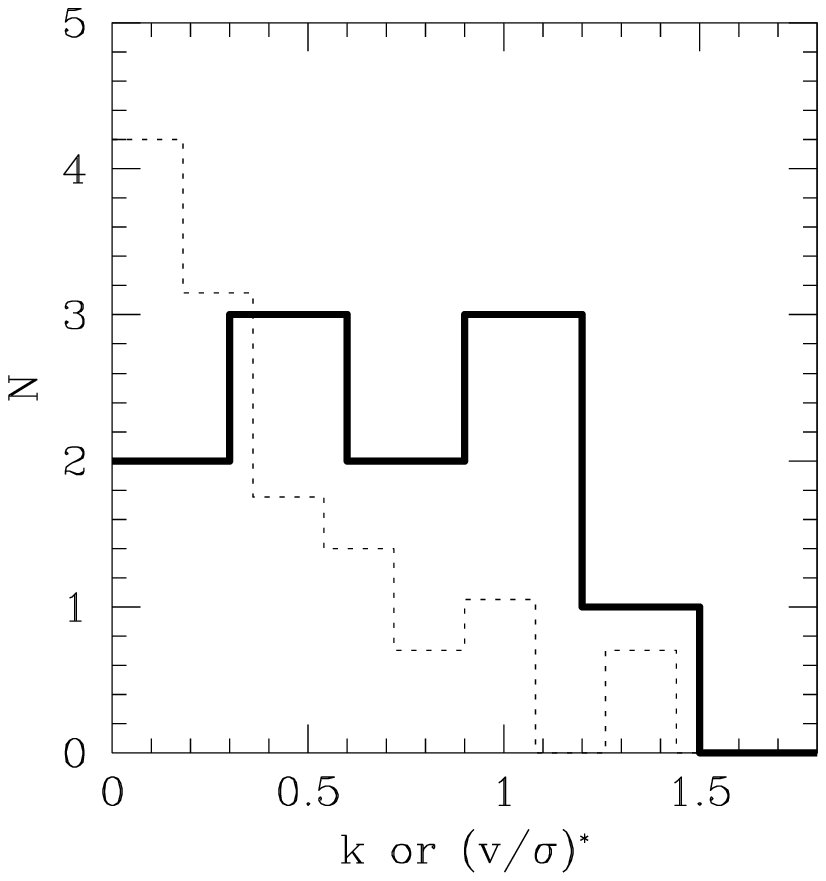}}
\figcaption{\figcapkhist}
\end{figure}
\fi

%%% END FIGURE %%%

\subsection{S0 Misclassification}
\label{ss:misclas}

One plausible explanation for the relatively high fraction of rapidly
rotating galaxies at $z \approx 0.5$ is that the fraction of
misclassified S0 galaxies among visually classified elliptical
galaxies is larger in intermediate-redshift clusters than it is
locally. This has in fact been suggested (e.g., Fabricant \etal 2000)
as a possible explanation for the low fraction of S0 galaxies found in
intermediate-redshift clusters (Dressler \etal 1997). The bulges of S0
and spiral galaxies typically have rotation rates consistent with
oblate isotropic rotator models (Davies \etal 1983). Any contamination
from disk light will further increase the observed
rotation. Therefore, one would normally expect to find S0 galaxies at
values of $k$ or $(v/\sigma)^{*}$ near or above unity.

Based on the statistics for the local comparison sample one would have
expected $0.19 \times 11 = 2.1$ galaxies in the distant galaxy
histogram in Figure~\ref{f:khist} with $k > 0.9$. Instead, 4 out of 11
are observed to have such high rotation rates. We excluded 4 galaxies
with $\epsilon < 0.10$ from the histogram. These are unlikely to be S0
galaxies, since the solid angle under which galaxies are viewed
pole-on is small. Hence, we estimate that 13\% (namely,
$(4-2.1)/(11+4)$) of the elliptical galaxies in these intermediate
redshift clusters might be misclassified S0 galaxies. The average of
the morphological percentages reported by Dressler \etal (1997) for
the three clusters studied here are E:S0:Spirals = 47:16:37. Upon
correction for the suggested misclassification of S0s as ellipticals
we obtain instead 41:22:37. Although correcting for misclassifications
increases the S0 fraction of these clusters, it remains much smaller
than the $\sim 60$\% found in local clusters (Dressler 1980). So there
appears to be a strong evolution in the real fraction of S0 galaxies
in clusters. Our data do not suggest that this can be attributed to
misclassification, although it should be kept in mind that the S0
fraction in intermediate redshift clusters might be higher than what
we just estimated. Not all S0 galaxies would necessarily be recognized
from their kinematics alone.

Our results with respect to potential S0 misclassification probably
depend on the sample selection. The three clusters in our sample were
selected based on their low fractions of visually-classified S0
galaxies (see Section~\ref{ss:sample}). On the one hand, this makes
these clusters good test cases: there are many visually-classified
elliptical galaxies that could potentially be misclassified S0
galaxies. On the other hand, if these clusters are truly poor in S0
galaxies, then they are not good test cases: in that case there simply
aren't many S0 galaxies that one could potentially
misclassify. Therefore, the results presented here may not generalize
to other clusters at similar redshift that have higher S0 fractions
based on visual classification.

\subsection{Redshift Evolution of Galaxy Rotational Support through Mergers}
\label{ss:rotevol}

An alternative interpretation for the difference between the
histograms in Figure~\ref{f:khist} is to assume that there is a true
redshift evolution in the internal dynamics of elliptical
galaxies. Since elliptical galaxies are collisionless systems, mergers
are required to significantly change their rotation rate. There are
different types of mergers and these can affect the evolution of the
observed rotation rate in different ways. We will discuss the relevant
types in turn. Independent of the type of merger, we are only
concerned here with mergers that happen between $z \approx 0.5$ and
$z=0$, i.e., in the last 5 Gyr.

Mergers between spiral galaxies form remnants that resemble elliptical
galaxies (e.g., Barnes 1988). The progenitors of such mergers are not
included in samples of elliptical galaxies at intermediate
redshift. However, the remnants of such mergers are included in local
samples of elliptical galaxies. Such mergers therefore contribute to
the so-called ``progenitor bias'' (e.g., van Dokkum \& Franx 2001).
Dissipationless and dissipational simulations show that the remnants
of disk-disk mergers have properties that are similar to local
low-luminosity ellipticals, which tend to be disky and flattened by
rotation (Naab \& Burkert 2003; Naab \etal 2006b). In contrast,
disk-disk mergers do not seem a plausible formation mechanism for the
most luminous ellipticals, since only equal-mass mergers with special
initial orientations can produce purely boxy anisotropic merger
remnants. So if it were possible to remove from the local sample in
Figure~\ref{f:khist} the ellipticals that formed from disk-disk
mergers since $z \approx 0.5$, then this would probably {\it decrease}
the relative proportion of rapidly-rotating ellipticals. Therefore,
progenitor bias due to disk-disk mergers cannot explain the evolution
seen in Figure~\ref{f:khist}. In fact, the progenitor bias is such
that the observed evolution is probably only a lower limit to the true
evolution.

The situation is different for mergers between elliptical galaxies.
The progenitors of such mergers are included in samples of elliptical
galaxies at intermediate redshift. Therefore, such mergers introduce
no progenitor bias (provided that luminosity-dependent sample
selection effects are properly understood). Simulations show that the
remnants of mergers between early-type galaxies have properties that
are similar to luminous ellipticals in the local universe (Naab \etal
2006a). In particular, the rotation rate of the remnants is generally
low, even when the progenitors were flattened by rotation. Because
elliptical galaxies have little gas, there is no formation of a gas
disk in the merger remnant that can form stars and boost the observed
rotation rate (in contrast to the situation for disk-disk mergers;
Naab \etal 2006b). Therefore, disspationless (``dry'') mergers between
elliptical galaxies can produce evolution in the elliptical galaxy
population towards lower rotation rates. This would be consistent with
the evolution seen in Figure~\ref{f:khist}.

The importance of mergers for the evolution seen in
Figure~\ref{f:khist} depends on the number of mergers of the relevant
types in the last 5 Gyr. There is observational evidence that merging
is indeed happening at an important rate in this time interval. Most
studies originally lumped various types of interactions and mergers
together (e.g., Le F\`evre \etal 2000; Patton \etal 2002; Conselice
\etal 2003; Lin \etal 2004; Bundy \etal 2004). However, some recent 
studies have aimed to separate mergers of different types to
specifically demonstrate the importance of dry merging (e.g., van
Dokkum 2005). Bell \etal (2006) used data for galaxies in the GEMS
survey to estimate that present-day spheroidal galaxies with $M_V <
-20.5$ have on average undergone between 0.5 and 2 dry mergers since
$z \approx 0.7$. Semi-analytical models of hierarchical structure
formation predict that in the mass-range of interest to the present
study, disk-disk mergers are about as common as dry mergers (Naab
\etal 2006a). Both types of mergers are therefore likely to have been 
numerous enough in the last 5 Gyr to have affected a significant
fraction of elliptical galaxies in the luminosity range covered by our
sample.

Most observational work of the merger rate as a function of redshift
has studied field galaxies. While these studies have shown that
mergers are important, this does not necessarily imply that our results
have been affected by them. This is because the frequency of merging
depends on environment. Our local comparison sample of elliptical
galaxy rotation rates (see Section~\ref{ss:localcomp}) is composed
mostly of galaxies in the field and in low-density cluster
environments such as Virgo. This range of environments is probably not
too dissimilar from that probed by field galaxy surveys out to
significant redshifts. This assumption implies that disk-disk mergers
in the last 5 Gyr have probably been numerous enough among our local
comparison sample to have affected the results in Figure~\ref{f:khist}
through the progenitor bias previously eluded to. 

Whether dry mergers have been numerous enough to decrease the rotation
rates of the elliptical galaxies in our intermediate-redshift sample
is a different issue. This is because those galaxies reside in dense
clusters. The traditional belief has been that galaxy interactions in
clusters are different from those in the field due to the large
relative velocities between galaxies. Galaxy transformations can then
occur due to the cumulative effect of frequent high speed galaxy
encounters (``galaxy harassment''; Moore \etal 1996). However,
calculations of merger rates (based mostly on somewhat idealized
simulations without a cosmological context; e.g., Makino \& Hut 1997)
show that true mergers may be rare (except for accretion events onto
the brightest cluster galaxy). However, recent observations have
challenged this paradigm. In particular, van Dokkum \etal (1999) and
Tran \etal (2005) found direct observational evidence for ongoing
mergers in the cluster MS 1054-03 at $z=0.83$. The true abundance of
dry mergers in clusters therefore remains unclear. However, it does
appear possible that dry mergers in clusters may have caused a decline
in the rotation rate of elliptical galaxies over the last 5 Gyr.

\subsection{Environmental Dependence and Cosmic Variance}
\label{ss:cosmic}

A final possible explanation for the difference between the rotation
rates of our sample galaxies at $z \approx 0.5$ and those of the local
comparison sample of Section~\ref{ss:localcomp} is that the average
rotation rate of elliptical galaxies may not be the same at different
locations in the Universe. There could be a dependence on environment
or there could be significant statistical fluctuations between widely
separated locations (cosmic variance). Our intermediate-redshift
sample consists of galaxies in dense clusters. In contrast, our local
comparison sample contains an inhomogeneous mixture of galaxies in
cluster, group and field environments. A dependence on environment
would therefore be particularly relevant for the comparison. However,
observations in the local universe have provided no evidence that such
a dependence exists. The $(v/\sigma)^*$ distributions of the rare
elliptical galaxies in very low-density environments are quite similar
to those for local samples biased towards high-density environments
(Hau \& Forbes 2006). Current cosmological models do not have the
resolution or detailed physics that would be needed to make confident
predictions for the size of either an environmental dependence or
cosmic variance in the elliptical galaxy rotation rate. A good way to
address this will therefore be to study larger samples of
intermediate-redshift galaxies in a wider range of locations and
environments.

\section{Summary and Discussion}
\label{s:conc}

We have presented detailed dynamical models for the spatially resolved
kinematics of 25 galaxies in the intermediate-redshift ($z
\approx 0.5$) clusters CL3C295, CL0016+16 and CL1601+42. The sample 
galaxies were selected to be bright enough for spectroscopy, and
visually classified from HST images as early type (and in most cases
elliptical) galaxies. The spectroscopic data were obtained with LRIS
on Keck as described in vDvdM06, with a $1.1''$ wide slit in seeing of
$0.71''$ to $0.91''$ FWHM. The data sample a region that extends
typically to $\sim 0.7''$ from the galaxy center ($\sim 4.3 \kpc$)
sampled at $0.215''$ per pixel.

To model the data we adopted an oblate axisymmetric luminosity density
of fixed axial ratio for each galaxy that fits the surface brightness
profile inferred from HST imaging. The inclination of the models was
chosen to provide the most likely intrinsic axial ratios in a
statistical sense. Edge-on models and highly-flattened models were
also explored, but the results do not depend strongly on the assumed
inclination. For each galaxy we solved the Jeans equations of
hydrostatic equilibrium under the assumption that the distribution
function has the two-integral form $f=f(E,L_z)$. The resulting model
predictions were projected along the line of sight and convolved with
the observational seeing, slit width, and binning along the
slit. Accounting for the observational characteristics is crucial for
a proper interpretation of resolved kinematical data obtained at these
redshifts, and in particular the rotation curves. When this is done,
the model predictions overall provide good fits to the kinematical
data.

Comparison of models and observations yields for each galaxy a
quantity $k$ that is a normalized measure of the rotation rate of the
galaxy. This quantity is similar to the quantity $(v/\sigma)^{*}$ that
is often used to measure the rotation rate of nearby galaxies (where
$k = (v/\sigma)^{*} = 1$ for an oblate isotropic rotator model). To
quantify the rotational properties of the sample galaxies we studied
how $k$ behaves as a function of galaxy magnitude $M_B$. We compared
this to the known correlations between $(v/\sigma)^{*}$ and $M_B$ for
elliptical galaxies in the local Universe. We find that the
intermediate-redshift ellipticals span a large range in $k$ from about
$0$ to $1.3$. This is not dissimilar from what is seen locally for
elliptical galaxies of similar luminosity. The four brightest galaxies
in the sample, which will have $M_B < -22$ even when evolved forward
in time to $z=0$, all rotate too slowly to account for their
flattening. This is consistent with what is seen in the local Universe
for galaxies of the same luminosity. Nonetheless, in a relative sense
there is a higher number of rapidly rotating galaxies in the
intermediate-redshift sample than in the local Universe (when
comparing galaxies of similar present-day luminosity).

The general similarity of the rotation rates of the sample galaxies to
those of local ellipticals suggests that the intermediate-redshift
galaxies have in large majority been correctly classified visually as
early-type galaxies. Nonetheless, a possible explanation for the
higher rotation rates of the galaxies at $z \approx 0.5$ is that the
fraction of misclassified S0 galaxies among visually classified
elliptical galaxies is larger in intermediate-redshift clusters than
it is locally. If this is indeed the cause of the observed effect,
then the S0 galaxy fraction in the target clusters may have been
underestimated by $\sim 6$\%. Even if this were the case, a strong
evolution in the S0 fraction of clusters continues to be implied
between the local Universe (where the fraction is $\sim 60$\%) and $z
\approx 0.5$ (where for the three clusters studied here the fraction
is only $\sim 22$\%, even after the listed upward correction). A
popular explanation for this evolution is that many local S0 galaxies
may have recently transformed from star-forming spiral
galaxies. However, this explanation is not without problems.
Substantial differences would be expected between the mean $M/L$ and
color of S0s and ellipticals, which are not observed (van Dokkum \&
Franx 2001). The contribution of misclassification to the observed
evolution should therefore remain a topic of investigation, in
particular since not all misclassified S0 galaxies would necessarily
be recognized from their kinematics alone.

Mergers can be another reason why there is a decrease in the observed
average rotation rate between $z \approx 0.5$ and the
present. Dissipationless (dry) mergers between early-type galaxies can
decrease the average rotation rate of the population, which is
qualitatively consistent with the direction of the observed
trend. However, it is unclear whether such mergers are numerous enough
in the high-velocity-dispersion environment of dense clusters to
reproduce the observed size of the trend. By contrast, disk-disk
mergers contribute to the comparison of our sample to local galaxies
only through the so-called ``progenitor bias''. Elliptical galaxies in
the local sample that formed from such mergers in the last 5 Gyr would
not be represented in the intermediate-redshift sample, which is
composed of visually-classified elliptical galaxies. This cannot
explain the observed trends though, because it increases rather than
decreases the relative number of rapidly rotating galaxies in the
local Universe. However, such mergers are numerous enough that their
progenitor bias must be taken into account in any quantitative
attempts to explain the observed trends.

With large aperture telescopes and multi-slit capabilities it is now
possible to study the resolved kinematics of large samples of galaxies
at intermediate redshifts. As we have shown here, detailed dynamical
modeling of such data has the potential to provide new insights into
the evolution of galaxies. This provides important advantages over
studies that are based solely on global quantities. For example,
changes in the rotation rates of galaxies as a function of redshift
can probe their merger history. It therefore seems useful to perform
more studies like the one presented here in the near future, to
explore wider ranges of redshifts, environments, and galaxy
types. Such studies would also explicitly constrain the contribution
to the observed differences in rotation rates between local elliptical
galaxies and those at intermediate redshifts due to potential
environmental dependencies and cosmic variance.

The models that we have constructed here also yield the mass-to-light
ratios $M/L$ for all the sample galaxies. These are used in Paper~II
to study the $M/L$ evolution of elliptical galaxies as a function of
redshift, which in turn constrains the formation ages of elliptical
galaxies. This complements and tests analyses based on Fundamental
Plane evolution, such as those presented in vDvdM06.

%%%%%%%%%%%%%%%
% Acknowledgments
%%%%%%%%%%%%%%%

\acknowledgments We thank Thorsten Naab for useful discussions. Part of this 
research was carried out at the Kavli Institute for Theoretical
Physics in Santa Barbara, supported in part by the National Science
Foundation under Grant No.~PHY99-07949.

%%%%%%%%%%%%%%%
% Appendix
%%%%%%%%%%%%%%%

% Uncomment if there are appendices
% \appendix

%%%%%%%%%%%%%%%
% Start references on a new page, unless we are in emulate mode
%%%%%%%%%%%%%%%

\ifemulate\else
\clearpage
\fi

%%%%%%%%%%%%%%%
% Use a small baselineskip for the references, unless in submission mode.
%%%%%%%%%%%%%%%

\ifsubmode\else\ifemulate\else
\baselineskip=10pt
\fi\fi

%%%%%%%%%%%%%%%
% Reference List
%%%%%%%%%%%%%%%

%%%%%%%%%%%%%%%
% Start captions + figures on a new page, unless we are in emulate mode
%%%%%%%%%%%%%%%

\ifemulate\else
\clearpage
\fi

%%%%%%%%%%%%%%%
% Change back to the regular baselineskip, if necessary
%%%%%%%%%%%%%%%

\ifsubmode\else\ifemulate\else
\baselineskip=14pt
\fi\fi

%%%%%%%%%%%%%%%
% Figures (in submission mode captions only, unless \printfigtrue)
%%%%%%%%%%%%%%%

\ifsubmode
\figcaption{\figcapphot}
\figcaption{\figcapkin}
\figcaption{\figcapklum}
\figcaption{\figcapkhist}
\clearpage
\else\printfigtrue\fi

\ifprintfig
\ifemulate\else

%%% FIGURE %%%

\clearpage
\begin{figure}
\epsfxsize=0.8\hsize
\centerline{\epsfbox{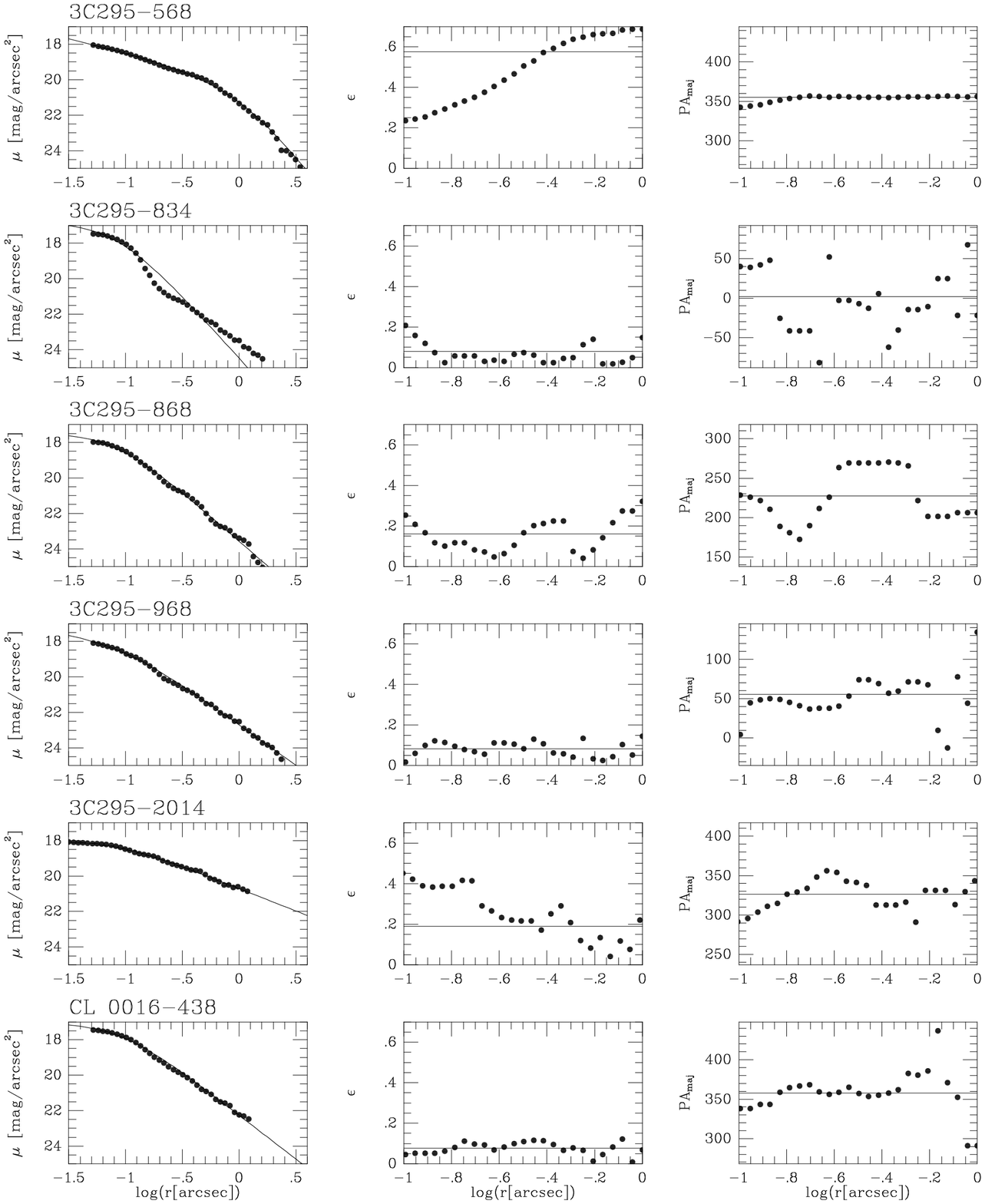}}
\ifsubmode
\vskip3.0truecm
\setcounter{figure}{0}
\addtocounter{figure}{1}
\centerline{Figure~\thefigure (part 1)}
\else
\figcaption{\figcapphot}
\addtocounter{figure}{-1}
\fi
\end{figure}

%%% FIGURE %%%

\clearpage
\begin{figure}
\epsfxsize=0.8\hsize
\centerline{\epsfbox{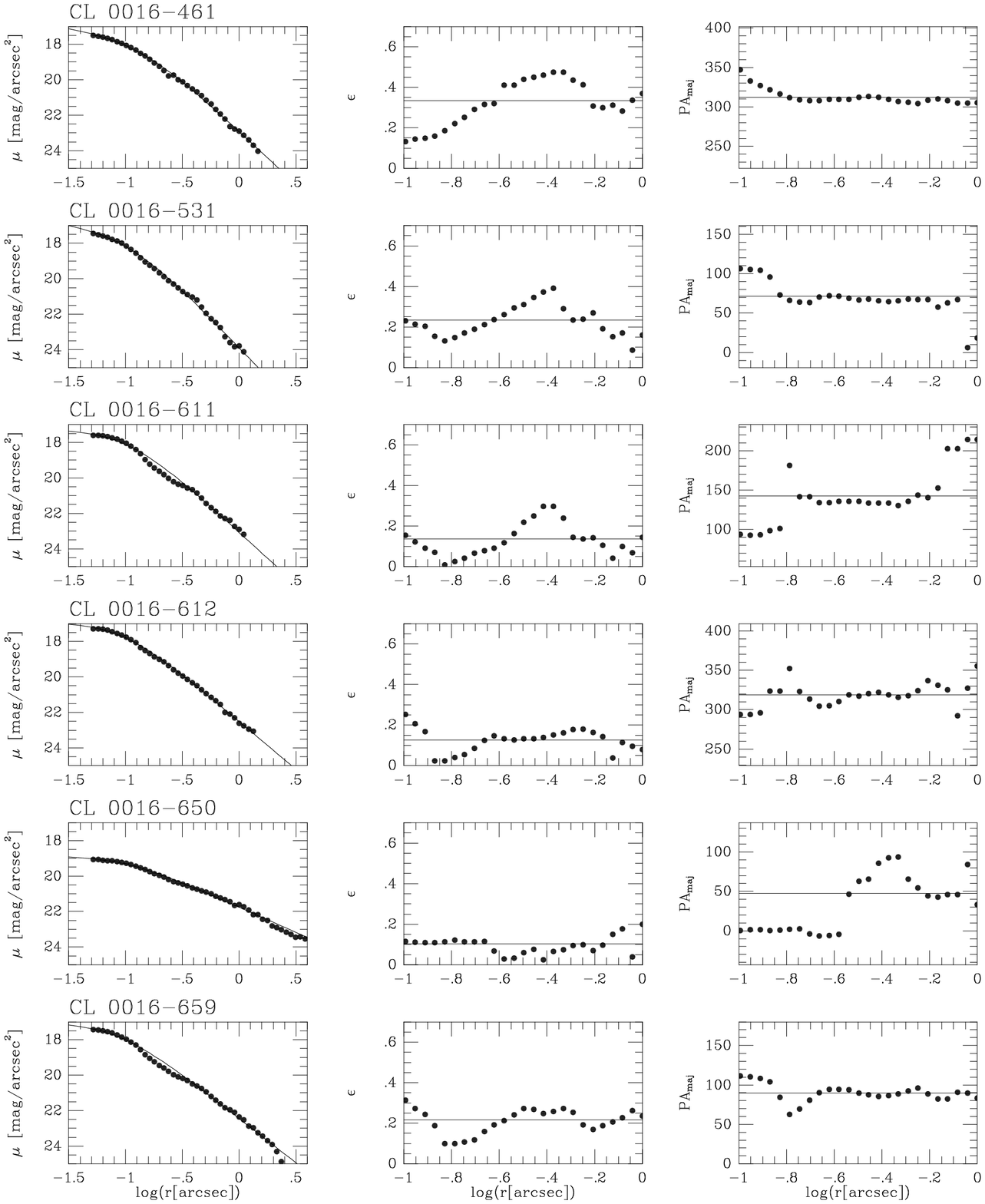}}
\ifsubmode
\vskip3.0truecm
\centerline{Figure~\thefigure (part 2)}
\else
\figcaption{(continued)}
\addtocounter{figure}{-1}
\fi
\end{figure}

%%% FIGURE %%%

\clearpage
\begin{figure}
\epsfxsize=0.8\hsize
\centerline{\epsfbox{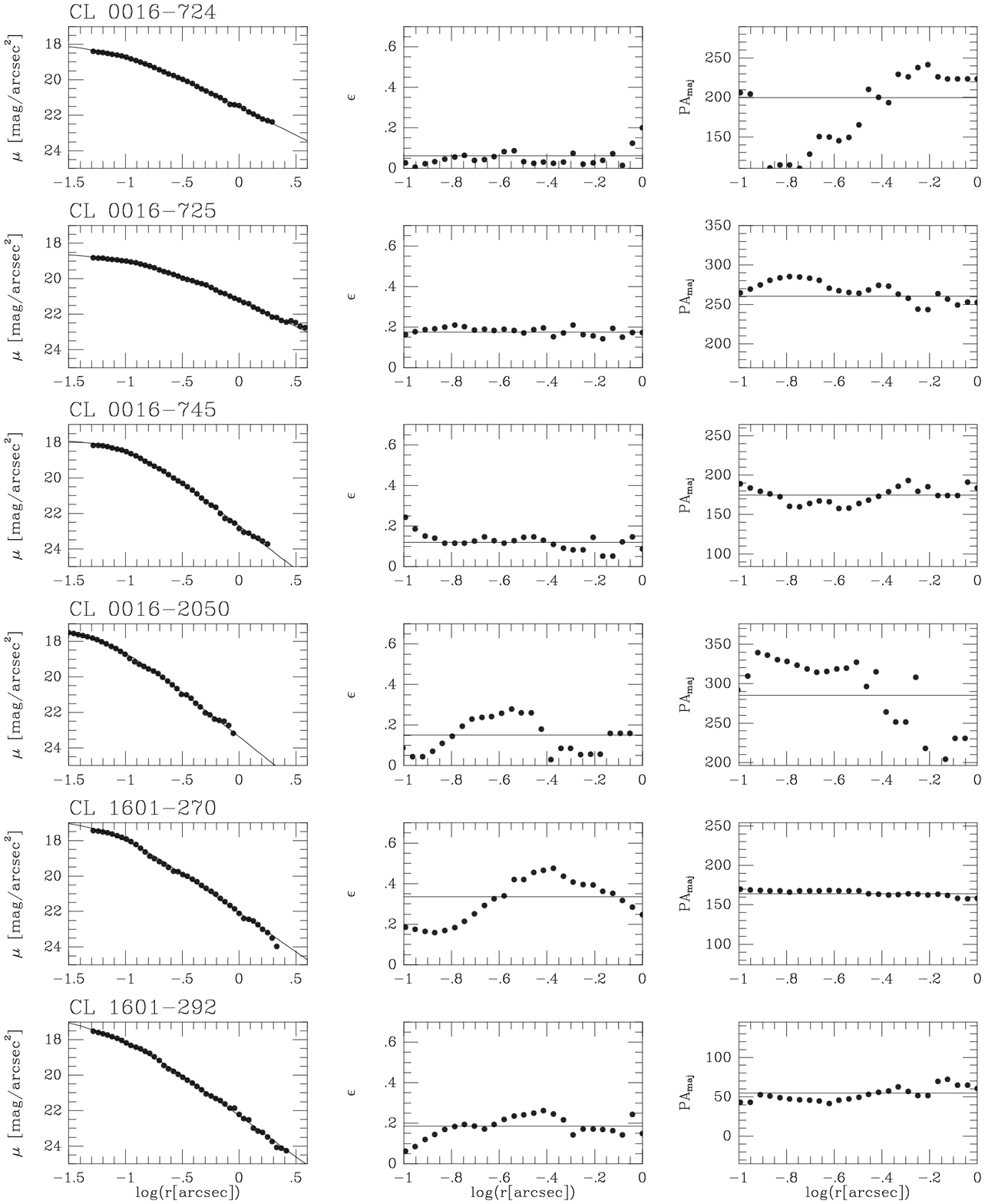}}
\ifsubmode
\vskip3.0truecm
\centerline{Figure~\thefigure (part 3)}
\else
\figcaption{(continued)}
\addtocounter{figure}{-1}
\fi
\end{figure}

%%% FIGURE %%%

\clearpage
\begin{figure}
\epsfxsize=0.8\hsize
\centerline{\epsfbox{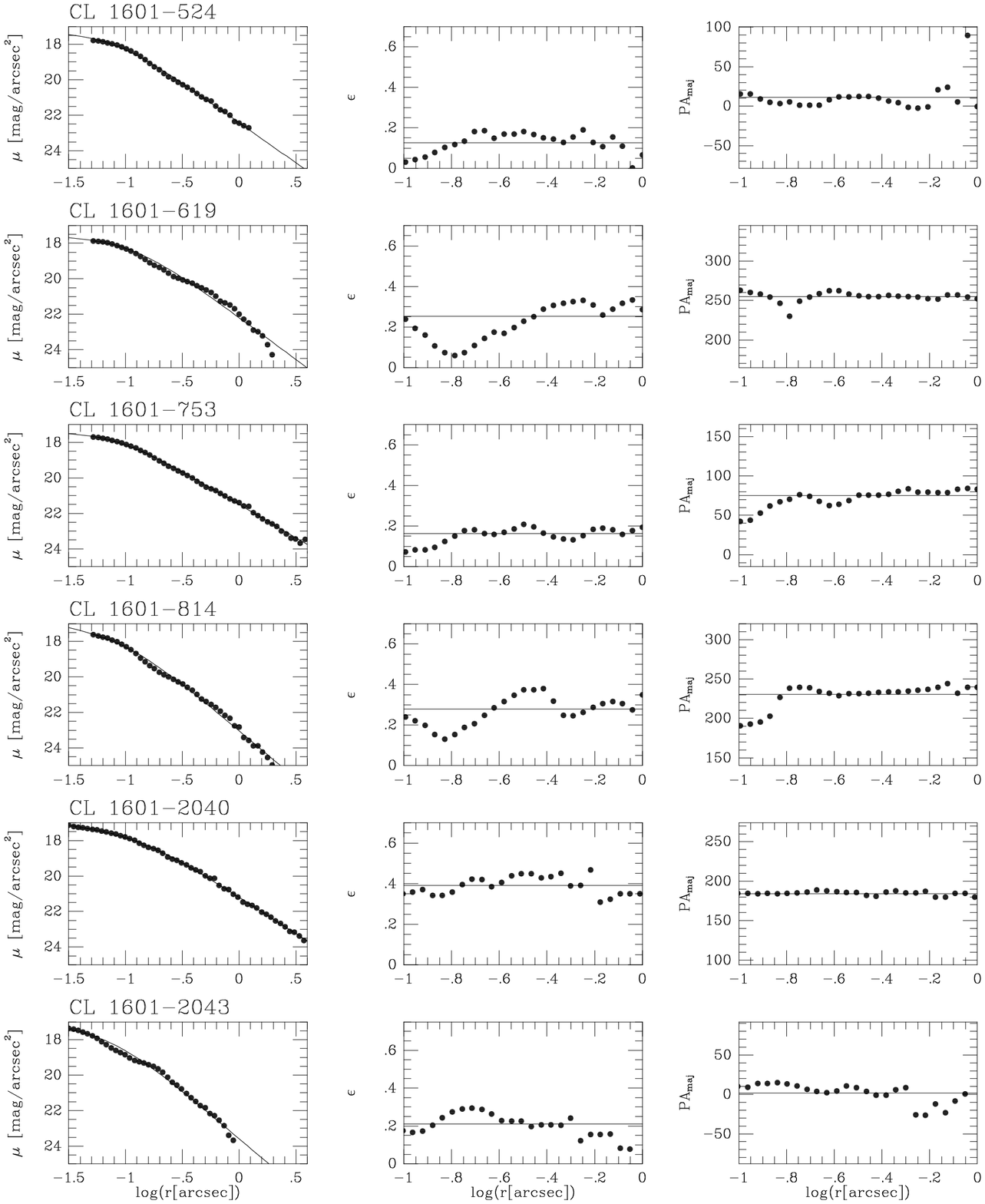}}
\ifsubmode
\vskip3.0truecm
\centerline{Figure~\thefigure (part 4)}
\else
\figcaption{(continued)}
\addtocounter{figure}{-1}
\fi
\end{figure}

%%% FIGURE %%%

\clearpage
\begin{figure}
\epsfxsize=0.8\hsize
\centerline{\epsfbox{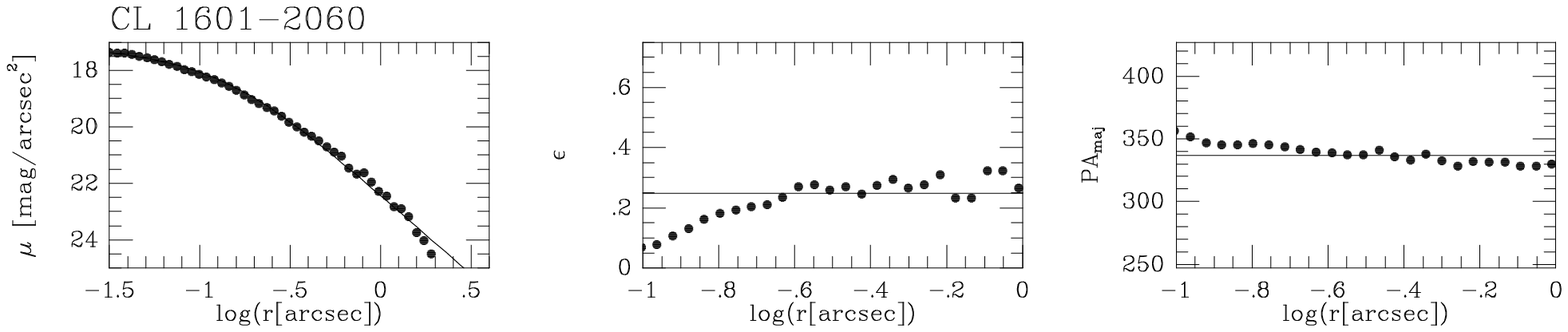}}
\ifsubmode
\vskip3.0truecm
\centerline{Figure~\thefigure (part 5)}
\else
\figcaption{(continued)}
\fi
\end{figure}

%%% FIGURE %%%

\clearpage
\begin{figure}
\epsfxsize=0.8\hsize
\centerline{\epsfbox{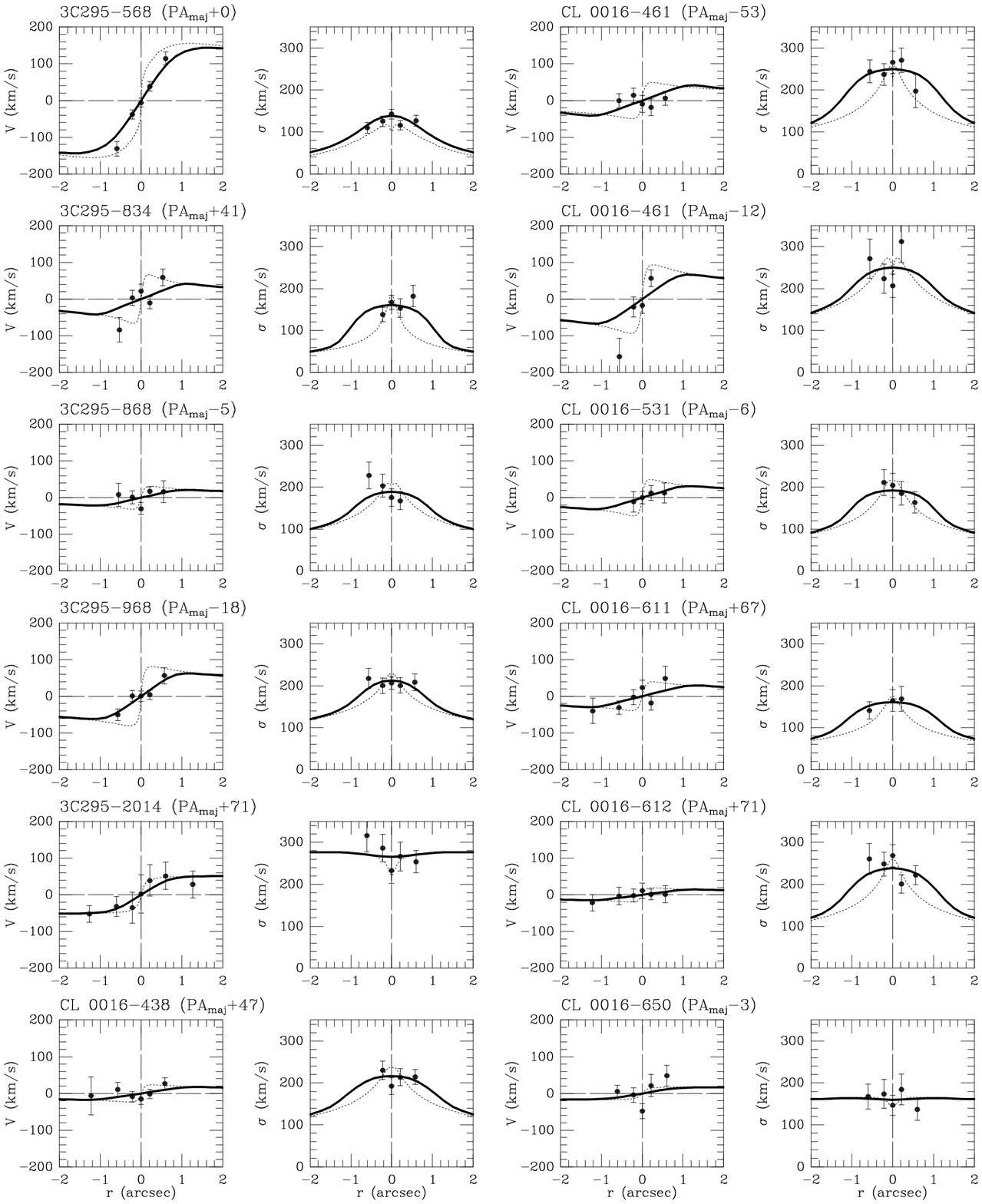}}
\ifsubmode
\vskip3.0truecm
\addtocounter{figure}{1}
\centerline{Figure~\thefigure (part 1)}
\else
\figcaption{\figcapkin}
\addtocounter{figure}{-1}
\fi
\end{figure}

%%% FIGURE %%%

\clearpage
\begin{figure}
\epsfxsize=0.8\hsize
\centerline{\epsfbox{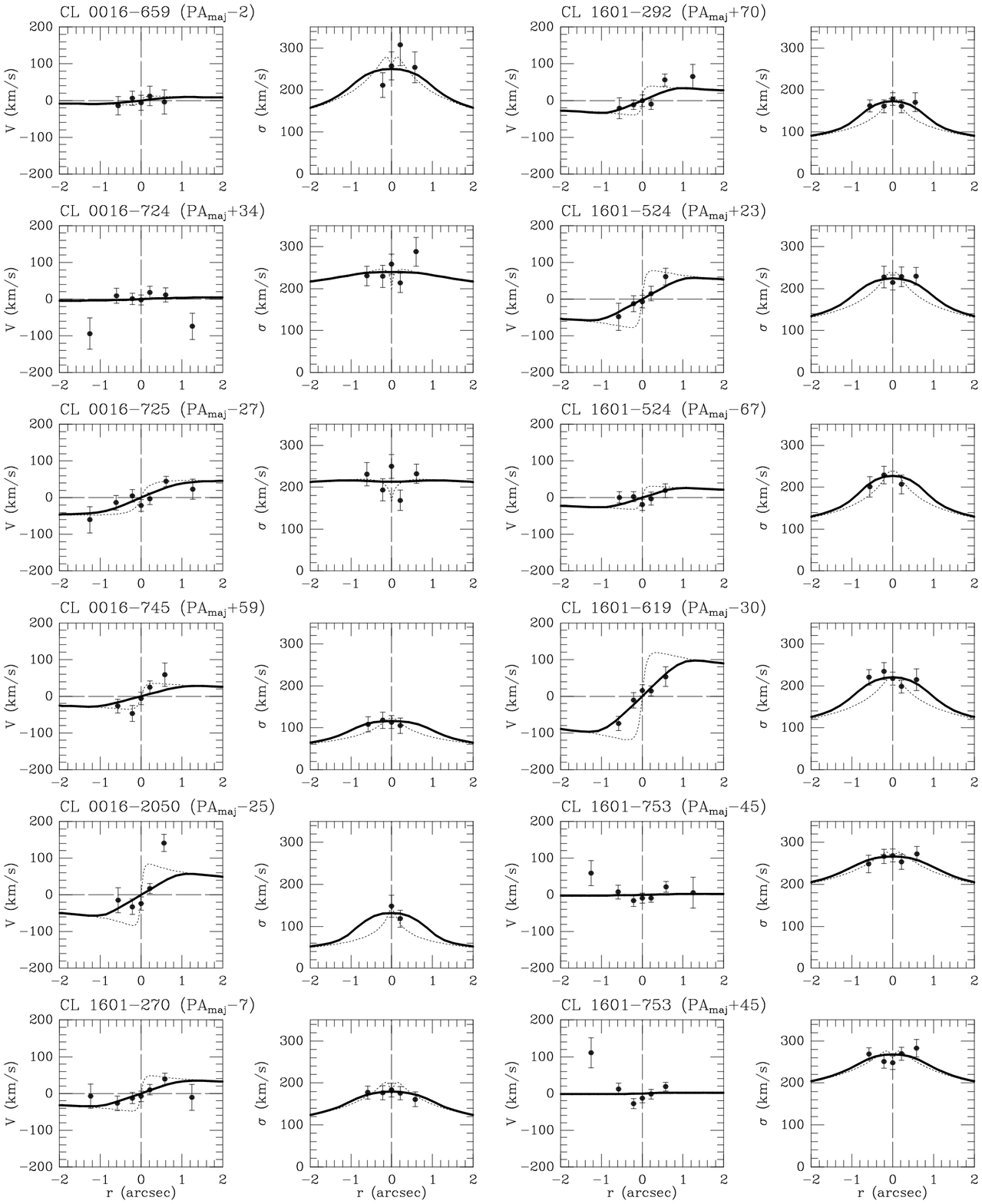}}
\ifsubmode
\vskip3.0truecm
\centerline{Figure~\thefigure (part 2)}
\else
\figcaption{(continued)}
\addtocounter{figure}{-1}
\fi
\end{figure}

%%% FIGURE %%%

\clearpage
\begin{figure}
\epsfxsize=0.39\hsize
\centerline{\epsfbox{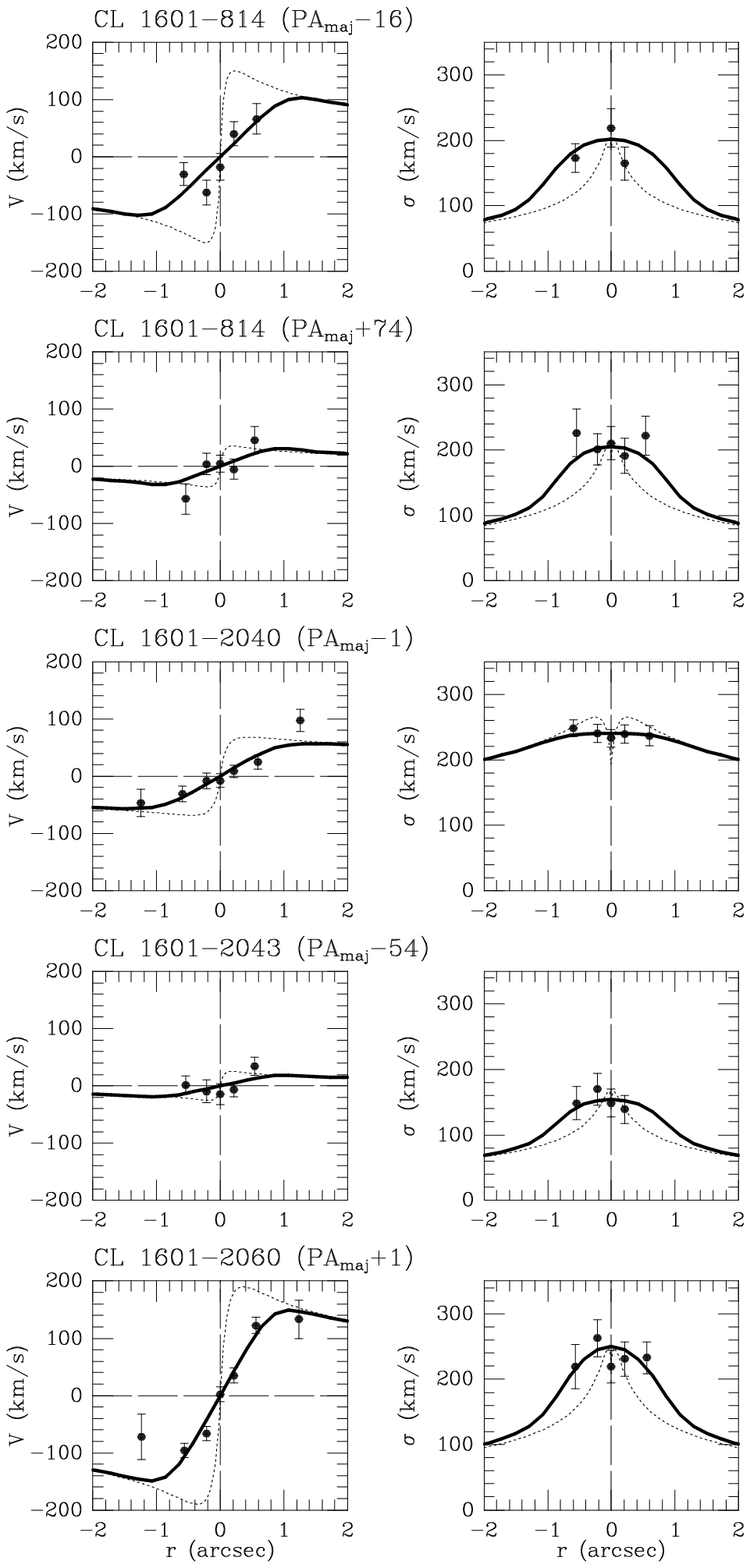}}
\ifsubmode
\vskip3.0truecm
\centerline{Figure~\thefigure (part 3)}
\else
\figcaption{(continued)}
\fi
\end{figure}

%%% FIGURE %%%

\clearpage
\begin{figure}
\epsfxsize=0.8\hsize
\centerline{\epsfbox{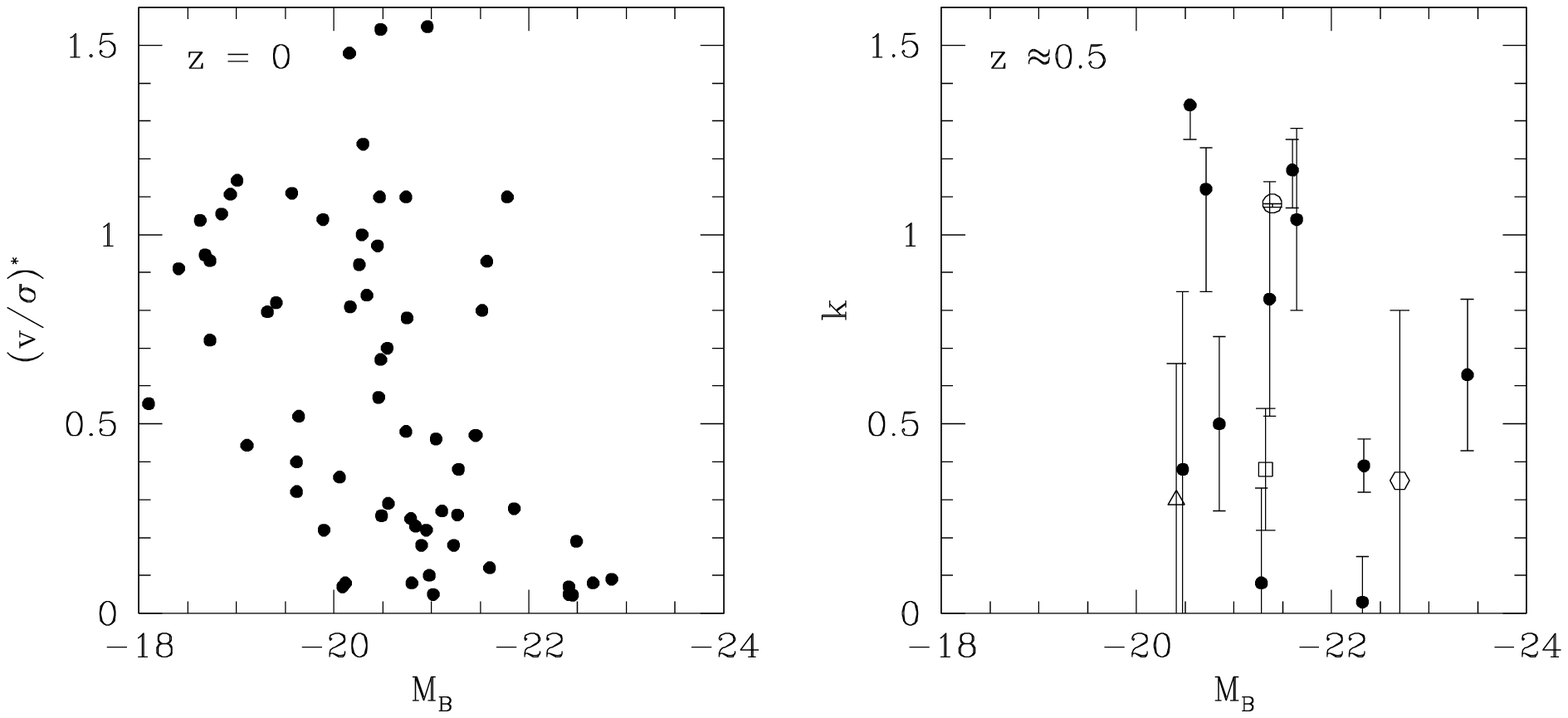}}
\ifsubmode
\vskip3.0truecm
\addtocounter{figure}{1}
\centerline{Figure~\thefigure}
\else
\figcaption{\figcapklum}
\fi
\end{figure}

%%% FIGURE %%%

\clearpage
\begin{figure}
\epsfxsize=0.4\hsize
\centerline{\epsfbox{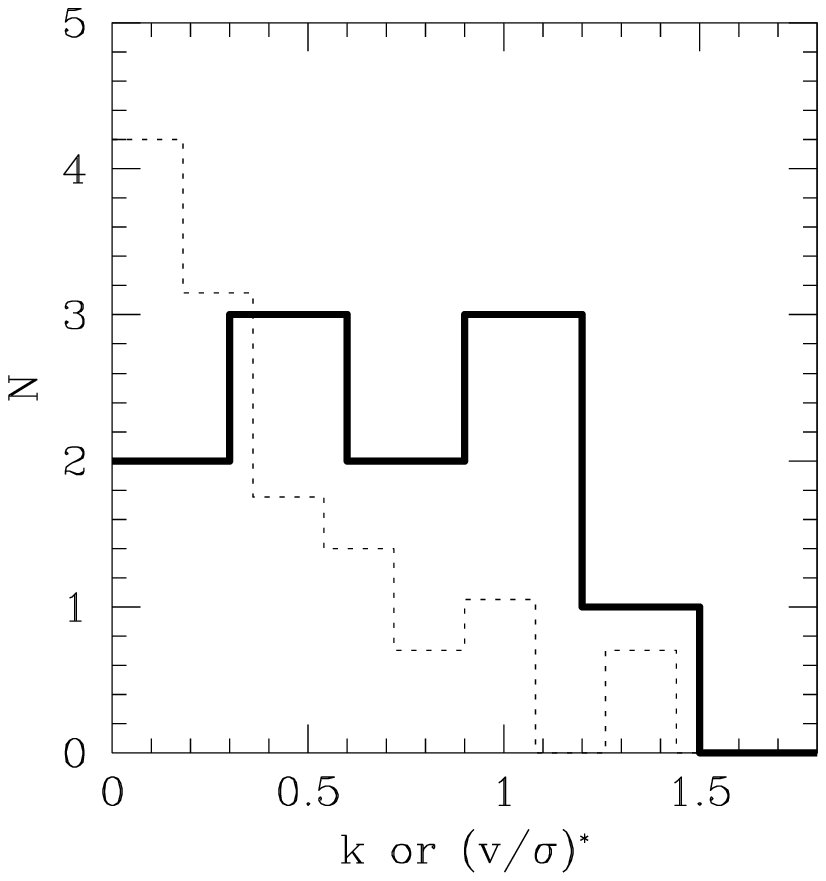}}
\ifsubmode
\vskip3.0truecm
\addtocounter{figure}{1}
\centerline{Figure~\thefigure}
\else
\figcaption{\figcapkhist}
\fi
\end{figure}

%%% END OF FIGURES %%%

\fi\fi

%%%%%%%%%%%%%%%
% Start tables on a new page, unless we are in emulate mode
%%%%%%%%%%%%%%%

\ifemulate\else
\clearpage
\fi

%%%%%%%%%%%%%%%
% Tables 
%%%%%%%%%%%%%%%

\ifsubmode\pagestyle{empty}\fi

\ifemulate\else
\begin{deluxetable}{llcrcccccccclcc}
\tabletypesize{\tiny}
\tablecontentsgalprop
\tablecomments{\tiny\tablecommgalprop}
\end{deluxetable}
\fi

%%% END OF TABLES %%%

%%%%%%%%%%%%%%%
% End of Document
%%%%%%%%%%%%%%%

\end{document}